\documentclass[sigconf]{acmart}

\setcopyright{none}
\renewcommand\footnotetextcopyrightpermission[1]{}

\AtBeginDocument{%
  }

\usepackage{booktabs}
\usepackage{multirow}
\usepackage{xcolor}
\usepackage{colortbl}
\usepackage[table]{xcolor}

\begin{document}

\title{HiEviDR-Bench: A Benchmark for Hierarchical Evidence Aggregation in Deep Research}

\author{Yubo Sun}
\authornote{Equal contribution.}
\affiliation{
  \institution{University of Chinese Academy \\ of Sciences \country{China}}
}
\email{boggysyb@gmail.com}

\author{Chunyi Peng}
\authornotemark[1]
\affiliation{
  \institution{Northeastern University}
\country{China}
}
\email{hm.cypeng@gmail.com}

\author{Yukun Yan}
\authornote{Corresponding author.}
\affiliation{
  \institution{Tsinghua University}
  \country{China}
}
\email{yanyk.thu@gmail.com}

\author{Zhenghao Liu}
\affiliation{
  \institution{Northeastern University}
  \country{China}
}
\email{liuzhenghao@mail.neu.edu.cn}

\author{Sen Mei}
\affiliation{
  \institution{Tsinghua University}
  \country{China}
}
\email{meisen2025@gmail.com }

\author{Bangrui Xu}
\affiliation{
  \institution{Shanghai Jiao Tong University}
  \country{China}
}
\email{dreameternal@sjtu.edu.cn}

\author{Xuanhe Zhou}
\affiliation{
  \institution{Shanghai Jiao Tong University}
  \country{China}
}
\email{zhouxuanhe@sjtu.edu.cn}

\author{Chi Chen}
\affiliation{
  \institution{Tsinghua University}
  \country{China}
}
\email{chenchithu@gmail.com}

\author{Maosong Sun}
\authornotemark[2]
\affiliation{
  \institution{Tsinghua University}
  \country{China}
}
\email{sms@tsinghua.edu.cn}


\begin{abstract}
Deep research requires models to retrieve, connect, and synthesize evidence from large-scale heterogeneous sources to answer complex queries and produce analytical reports. Existing benchmarks mainly evaluate final outcomes, such as answer correctness, report quality, or citation alignment, while providing limited visibility into whether evidence is correctly selected, linked, and aggregated into supported claims and conclusions. To address this gap, we introduce \textbf{HiEviDR-Bench}, a benchmark for evaluating \textbf{Hi}erarchical \textbf{Evi}dence aggregation in \textbf{D}eep \textbf{R}esearch. HiEviDR-Bench covers open-domain and academic-domain settings under both text-only and multimodal conditions, and represents each instance with an explicit evidence graph that captures evidence selection, cross-source linking, and aggregation from evidence to intermediate claims and final conclusions. Based on this formulation, we develop a traceability-oriented evaluation framework with five dimensions—report quality, evidence traceability, citation accuracy, claim verification, and answer correctness—together with a progressive gating mechanism for fine-grained error localization. HiEviDR-Bench contains 2,000 human-validated questions with evidence graphs across multiple difficulty levels. Experiments on 16 representative multimodal large language models show that, although many systems achieve strong report quality, their performance drops markedly on citation accuracy, claim construction, and answer correctness. Further analysis shows that the main bottlenecks lie in evidence identification and intermediate claim construction, revealing that the strong surface-level report quality does not necessarily imply grounded multi-stage reasoning on our benchmark.
All codes and datasets are available at {\color{blue}\url{https://ai9stars.github.io/HiEviDR-Bench.github.io}}.
\end{abstract}




\maketitle

\section{Introduction}


\begin{figure*}[t!]
  \centering
  \includegraphics[width=1.0\textwidth]{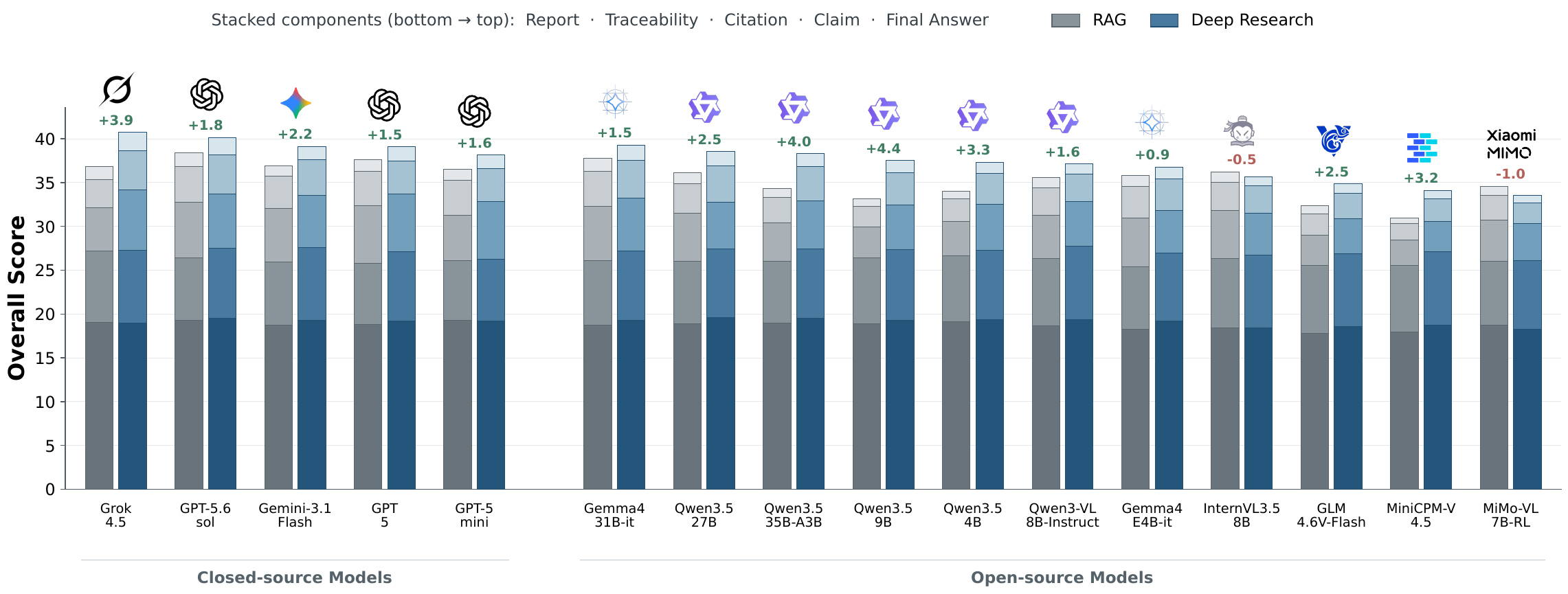}  
\caption{Stage-wise score of HiEviDR-Bench, ranked by overall score(0-100).}
  \label{fig:intro}
\end{figure*}

Deep research refers to the task of retrieving, synthesizing, and aggregating information from large-scale heterogeneous sources to produce comprehensive analytical reports in response to complex queries~\cite{huang2025deep, du2025deepresearch, liu2026knowledge}.
Unlike conventional question answering that targets short factual responses, it requires multi-step retrieval, cross-source evidence linking, and structured long-form generation grounded in verifiable evidence chains~\cite{zhang2025deep, hu2026sage}.
Recent advances in Large Language Models (LLMs)~\cite{yang2025qwen3, zeng2026glm, seed2025seed2} have given rise to deep research methods that can autonomously decompose queries, perform iterative retrieval, and generate citation-grounded reports resembling human research outputs~\cite{feng2026idrbench}.

Despite this progress, current evaluation of deep research remains limited, where most existing benchmarks adopt an outcome-centric paradigm, assessing only the final report through LLM-as-judge scoring or reference-based matching while treating the evidence aggregation process as a black box~\cite{wei2025browsecomp, li2025reportbench, wang2025liveresearchbench}.
Such evaluation cannot diagnose the source of errors: retrieval failure, reasoning mistakes, or citation inaccuracy, nor does it provide structured metrics for whether retrieved evidence is correctly organized into coherent intermediate claims~\cite{sorodoc2025garage}.
Meanwhile, multimodal scenarios involving tables, charts, and figures as evidence remain largely unexplored~\cite{huang2026mmdeepresearch, zeng2026vision}.
Without traceability into the hierarchical evidence chain, existing evaluation can reveal that a method fails but not why~\cite{    ye2026miroeval, zhang2025far}.

To address these limitations, we introduce \textbf{HiEviDR-Bench}, a benchmark designed to evaluate hierarchical evidence aggregation in deep research with full traceability.
For each query, we construct a hierarchical evidence graph, a directed acyclic graph with three levels: evidence nodes capturing atomic supporting facts including both textual and visual items, claim nodes representing intermediate reasoning steps, and a conclusion node encoding the final answer, making every link in the reasoning chain explicit and inspectable.
On top of this graph, we develop a traceability-oriented evaluation framework comprising five complementary dimensions, namely report quality, evidence traceability, citation accuracy, claim verification, and answer correctness, coupled with a progressive gating mechanism that checks whether each stage is faithfully completed before proceeding to the next.
The benchmark spans open-domain and academic domain-specific evaluation, yielding 2,000 human-validated questions with evidence graphs across text-only and multimodal settings at three difficulty levels.

Extensive experiments across 16 competitive multimodal large language models under both the RAG and deep research paradigms reveal a consistent gap between report quality and faithful evidence aggregation. Although most models can produce fluent and well-structured reports, their performance drops markedly in citation accuracy, claim construction, and answer correctness, indicating that high-quality reports do not necessarily reflect grounded multi-stage reasoning. 
Further progressive gating analysis shows that the main bottlenecks arise earlier in the pipeline, especially in evidence identification and intermediate claim construction, with answer-stage gate pass rates ranging from only 3.80\% to 11.50\%. Taken together, our results show that the central limitation of existing deep research systems lies in evidence composition and claim-level reasoning, rather than in report fluency alone.


\section{Related Work}

Recent advances in Large Language Models (LLMs) have substantially expanded the scope of automated knowledge-intensive tasks, moving from short-form question answering to long-form, evidence-grounded report generation. Retrieval-augmented generation (RAG) methods enable models to access external knowledge and produce contextually grounded responses~\cite{nakano2021webgpt,sun2025visrag, xiong2026lang2act, peng2026mixture}, while more recent Deep Research Agents (DRAs) further extend this paradigm by introducing iterative retrieval, query decomposition, multi-document exploration, and multi-step synthesis~\cite{xu2025comprehensive,shao2024assisting,du2025deepresearch,xiong2026paperscope}. As a result, deep research has emerged as a representative task in which models must assemble heterogeneous evidence, connect cross-source information, and generate citation-grounded analytical reports that resemble human research workflows~\cite{venkit2025deeptrace}.

Despite this progress, existing evaluation of deep research remains largely outcome-centric. Most benchmarks assess primarily final outputs, such as answer correctness, report quality, or citation quality, while treating much of the evidence aggregation process as a black box~\cite{huang2026mmdeepresearch,wang2025liveresearchbench}. Such evaluation can reveal whether a system succeeds or fails, but provides little insight into why it fails, for example, whether errors arise from evidence recall, evidence organization, citation attribution, or intermediate reasoning. Recent studies have begun to move toward more diagnostic evaluation by examining intermediate reasoning, grounding signals, retrieval stages, and evidence-linked reasoning chains~\cite{chen2026trace,xiao2025graphrag,ren2026sin}. Nevertheless, current benchmarks still lack an explicit framework for evaluating whether retrieved evidence is correctly organized into coherent intermediate claims and final conclusions. This limitation motivates the need for a benchmark that evaluates deep research not only by final outputs but also by the hierarchical evidence chain underlying them.

Another important limitation of existing benchmarks is their predominantly text-centric formulation. Real-world deep research often involves heterogeneous evidence sources, including text, tables, charts, and figures, and requires models to integrate them into a unified reasoning process. However, most current benchmarks either focus on textual outputs or assess multimodal understanding in relatively narrow settings~\cite{zeng2026vision,chen2025browsecomp}. Although tasks such as MMQA~\cite{yu2025mramg}, DocVQA~\cite{tito2023hierarchical}, MM-BrowseComp~\cite{li2025reportbench}, and BrowseComp-$V^3$~\cite{zhang2026browsecomp} have advanced the evaluation of multimodal retrieval, browsing, and text--vision integration, they are not specifically designed to assess hierarchical evidence aggregation in long-form deep research. This leaves open the question of how well current systems can trace, organize, and compose multimodal evidence across multiple reasoning stages. Therefore, a benchmark for deep research should go beyond end-to-end output evaluation and explicitly measure multimodal evidence aggregation with process-level traceability.

\section{HiEviDR-Bench}
\label{sec:bench}

\begin{figure*}[t!]
  \centering
  \includegraphics[width=1.0\textwidth]{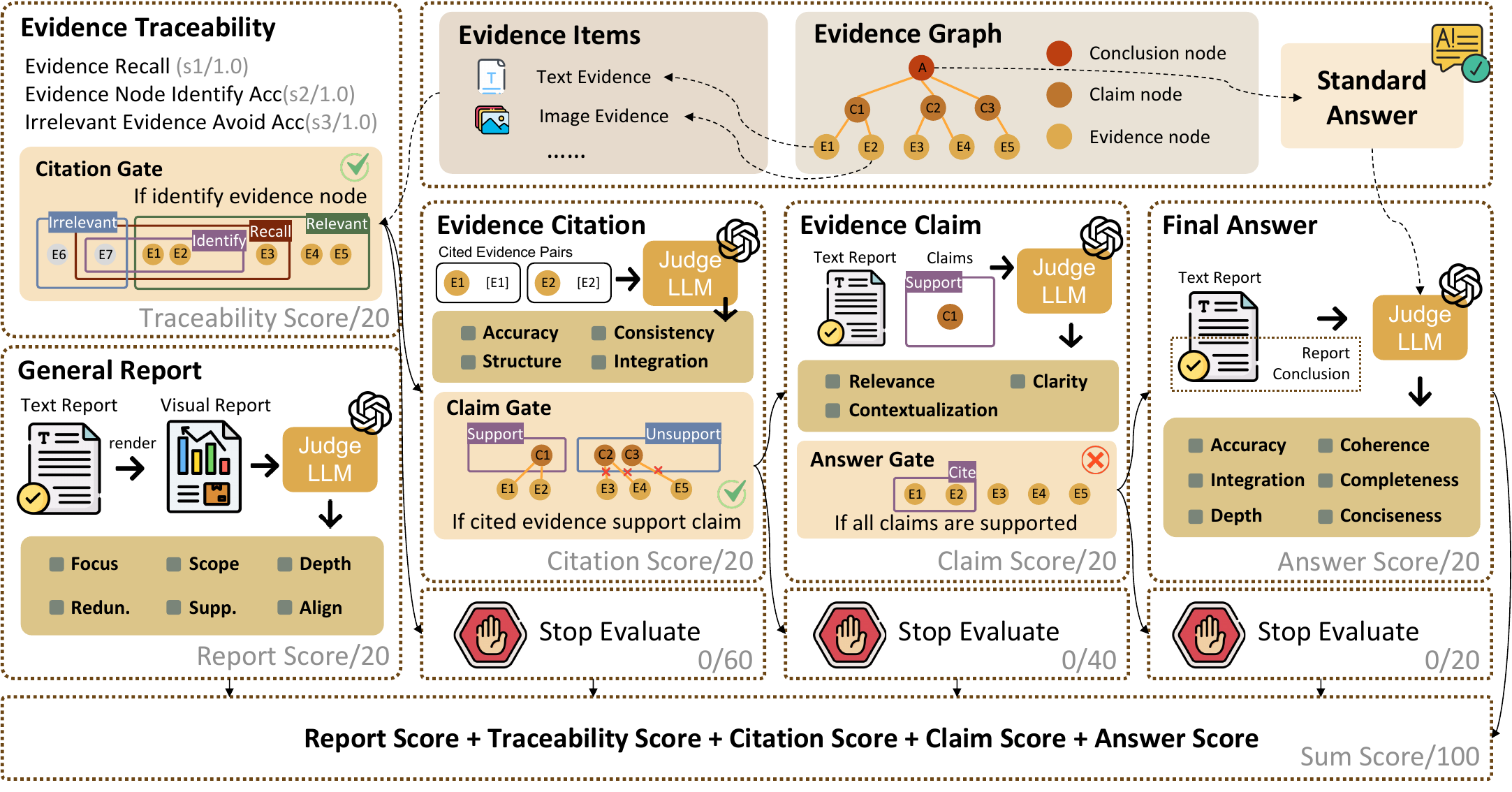}  
\caption{Overview of the evaluation framework in HiEviDR-Bench. Each question is paired with heterogeneous evidence items, an evidence graph, and a standard answer, enabling five-dimensional evaluation of a generated report: Report, Traceability, Citation, Claim, and Answer. A progressive gating mechanism is applied to the last three stages to ensure faithful evidence aggregation and grounded answer generation.}
  \label{fig:evaluation}
\end{figure*}

In this section, we present HiEviDR-Bench, a benchmark designed for evaluating hierarchical evidence aggregation in deep research.
We begin by formalizing the multimodal deep research task and contrasting our traceability-oriented evaluation philosophy with existing outcome-oriented approaches in Sec.~\ref{sec:task_def}.
We then introduce the multi-dimensional evaluation framework built upon the evidence graph, covering both multimodal report assessment and evidence-grounded traceability evaluation in Sec.~\ref{sec:eval_dim}.
Finally, we describe the data curation pipeline, detailing how the evidence graph and question--answer pairs are constructed in Sec.~\ref{sec:data_curation}.

\subsection{Task Formulation of Deep Research}
\label{sec:task_def}

Existing deep research benchmarks~\cite{huang2026mmdeepresearch} typically formulate evaluation as an end-to-end generation task.
Given a complex query $q$ and a large multimodal corpus $\mathcal{C}$, a deep research method first retrieves a set of relevant evidence items
\begin{equation}
  E = \text{Retrieve}(q, \mathcal{C}),
  \label{eq:retrieve}
\end{equation}
where each evidence item may be textual, visual, or document-grounded. The method then aggregates the retrieved evidence and produces a long-form research report
\begin{equation}
  R = \mathcal{M}(q, E),
  \label{eq:generate}
\end{equation}
where $R$ denotes the final output rather than a short-form answer.
A practical characteristic of this task is that multimodality may appear on both sides: the query may include mixed visual-textual context $q = \{q^{\text{text}}, q^{\text{img}}\}$, and the report may likewise contain both textual and visual content $R = \{R^{\text{text}}, R^{\text{img}}\}$, where $R^{\text{img}}$ includes images, figures, or tables used to support the final presentation.

These benchmarks then evaluate the generated report through a holistic quality assessment. Given the report $R$ and a gold reference answer $A^{*}$, the evaluation score $S$ is computed as
\begin{equation}
  S = \text{Judge}(R, A^{*}, q),
  \label{eq:existing_eval}
\end{equation}
where the judge directly compares the final report against the reference, assessing overall correctness, fluency, or citation alignment.
This formulation treats the reasoning pipeline as a black box~\cite{huang2026mmdeepresearch,wang2025liveresearchbench}: a method may achieve high scores under Eq.~\ref{eq:existing_eval} even when key evidence is missed or claims lack adequate support, as long as the final output appears reasonable.

In contrast, HiEviDR-Bench formulates deep research evaluation as a hierarchical evidence aggregation task where every conclusion is traceable to its supporting evidence.
Rather than assessing only the final report, our benchmark annotates a structured evidence graph $\mathcal{G}$ for each question.
Formally, we define the evidence graph as a directed acyclic graph
\begin{equation}
  \mathcal{G} = (\mathcal{N}, \mathcal{E}),
  \label{eq:graph}
\end{equation}
where the node set $\mathcal{N} = \mathcal{N}_{\text{e}} \cup \mathcal{N}_{\text{c}} \cup \{n_{\text{conclusion}}\}$ consists of three types: evidence nodes $\mathcal{N}_{\text{e}}$ representing atomic evidence items retrieved from the corpus, claim nodes $\mathcal{N}_{\text{c}}$ representing intermediate reasoning statements, and a single conclusion node $n_{\text{conclusion}}$ denoting the final answer.
The edge set $\mathcal{E} \subseteq \mathcal{N} \times \mathcal{N}$ captures the support relations among nodes: an edge $(n_i, n_j) \in \mathcal{E}$ indicates that node $n_i$ provides direct evidential or logical support for node $n_j$.

Under this formulation, the evaluation of a model-generated report $R$ is no longer a single-step correctness check as in Eq.~\ref{eq:existing_eval}, but instead a traceability-oriented measurement grounded in $\mathcal{G}$.
Given the evidence graph, the total evaluation score is decomposed into five complementary dimensions, each capturing a distinct stage of the evidence-to-answer chain:
\begin{equation}
  S = S_{\text{report}} + S_{\text{trace}} + S_{\text{citation}} + S_{\text{claim}} + S_{\text{answer}},
  \label{eq:total}
\end{equation}
where each component is scored in $[0, 20]$, yielding a total out of 100.

\begin{table}[t]
\centering
\caption{Corpus statistics of HiEviDR-Bench.}
\label{tab:corpus_statistics}
\begin{tabular}{lcc}
\toprule
\textbf{Domain} & \textbf{\#Text Items} & \textbf{\#Image Items} \\
\midrule
Wiki  & 30,463,973 & 6,676,595 \\
Arxiv & 124,650    & 42,193   \\
\bottomrule
\end{tabular}
\end{table}
\subsection{Evaluation Dimensions}
\label{sec:eval_dim}

A central design principle of HiEviDR-Bench is that every aspect of the final score should be traceable to concrete evidence grounding, rather than relying on holistic quality judgments that conflate surface fluency with faithful reasoning~\cite{xiao2025graphrag}.
As illustrated in Figure~\ref{fig:evaluation}, each question is paired with heterogeneous evidence items, an evidence graph $\mathcal{G}$, and a standard answer.
Given a model-generated report, we decompose its quality into five scores, each in $[0,20]$: Report Score (multimodal report assessment), Evidence Traceability Score, Citation Score, Claim Score, and Answer Score (evidence traceability evaluation), where the last three are further constrained by a progressive gating mechanism.

\textbf{Report Score.}
Since the report may contain visual elements, we render it into a visual form and provide it to a Judge-Model.
Let $\mathcal{K}_{r}$ denote the set of six report-quality dimensions: Focus (Topical Focus and Coherence), Scope (Knowledge Breadth and Coverage), Depth (Depth of Analysis and Reasoning), Redun. (Redundancy and Information Density), Supp. (Grounding and Evidential Support), and Align (Visual--Textual Alignment).

Let $r_k \in [1,5]$ be the Judge-Model score for dimension $k \in \mathcal{K}_{r}$.
The Report Score is
\begin{equation}
S_{\text{report}}
=
20 \cdot \frac{1}{5|\mathcal{K}_{r}|}\sum_{k \in \mathcal{K}_{r}} \frac{r_k-1}{4},
\label{eq:report}
\end{equation}
where $|\mathcal{K}_{r}|$ denotes the number of dimensions in $\mathcal{K}_{r}$.

\textbf{Evidence Traceability Score.}
Before evaluating higher-level reasoning, we first assess whether a model can retrieve relevant evidence, identify key evidence nodes, and suppress truly irrelevant evidence.
This stage consists of three complementary metrics, each in $[0,1]$: Evidence Recall ($s_1$), Evidence Node Identify Accuracy ($s_2$), and Irrelevant Evidence Avoid Accuracy ($s_3$).

Evidence Recall measures the proportion of gold key evidence successfully retrieved by the model:
\begin{equation}
s_1
=
\mathrm{Recall}(E_{\text{key}}^{\text{pred}},E_{\text{key}}^{\text{gold}})
=
\frac{|E_{\text{key}}^{\text{pred}}\cap E_{\text{key}}^{\text{gold}}|}
{|E_{\text{key}}^{\text{gold}}|},
\label{eq:s1}
\end{equation}
where $E_{\text{key}}^{\text{gold}}$ denotes the gold key-evidence set and
$E_{\text{key}}^{\text{pred}}$ is the model-predicted key-evidence set.

Evidence Node Identify Accuracy further evaluates whether the retrieved key evidence is correctly utilized in the model’s final reasoning process:
\begin{equation}
s_2
=
\frac{|E_{\text{cite}}^{\text{pred}}\cap E_{\text{key}}^{\text{gold}}|}
{|E_{\text{key}}^{\text{gold}}|},
\label{eq:s2}
\end{equation}
where $E_{\text{cite}}^{\text{pred}}$ denotes the evidence nodes explicitly cited by the model.
This metric measures the proportion of gold key evidence that is correctly identified by the model.

Irrelevant Evidence Avoid Accuracy measures the model's ability to filter out truly irrelevant evidence during reasoning.
To avoid penalizing evidence that is semantically related to the query but absent from the annotated graph, we first compute the TF-IDF similarity between each recalled evidence item and the input question.
Only evidence whose similarity is lower than a predefined threshold $\tau$ is regarded as truly irrelevant.
Let $E_{\text{irr}}^{\text{gold}}$ denote this filtered irrelevant evidence set and
$E_{\text{avoid}}^{\text{pred}}$ denote the subset that the model successfully excludes from its generated report.
The metric is computed as
\begin{equation}
s_3
=
\frac{|E_{\text{avoid}}^{\text{pred}}\cap E_{\text{irr}}^{\text{gold}}|}
{|E_{\text{irr}}^{\text{gold}}|},
\label{eq:s3}
\end{equation}
where $E_{\text{irr}}^{\text{gold}}$ contains only evidence items whose TF-IDF similarity to the question is below the threshold $\tau$.
This filtering reduces false penalties caused by semantically relevant evidence that is not included in the annotated evidence graph.

The final Evidence Traceability Score aggregates the three complementary metrics:
\begin{equation}
S_{\text{trace}}
=
20\cdot
\frac{2s_1+s_2+s_3}{4},
\label{eq:trace}
\end{equation}

Together, these three metrics provide a fine-grained evaluation of evidence retrieval, key-node identification, and irrelevant-evidence filtering before higher-level evidence aggregation takes place.

\textbf{Progressive Gating Mechanism.}
The remaining three dimensions: Citation Score, Claim Score, and Answer Score, are evaluated in a cascaded fashion governed by a progressive gating mechanism, where a report must satisfy the grounding requirement of the preceding stage before receiving credit in the next one; failure at any gate zeroes out all subsequent stages.

Concretely, the Citation Gate activates Citation Score only when the model cites relevant evidence, corresponding to the activation of at least one evidence node in $\mathcal{N}_{\text{e}}$.
The Claim Gate activates Claim Score only when the cited evidence adequately supports the corresponding claim.
The Answer Gate activates Answer Score only when all required claims are supported, interpretable as the full activation of $\mathcal{G}$ up to the conclusion node $n_{\text{conclusion}}$.

\begin{figure*}[t!]
  \centering
  \includegraphics[width=1.0\textwidth]{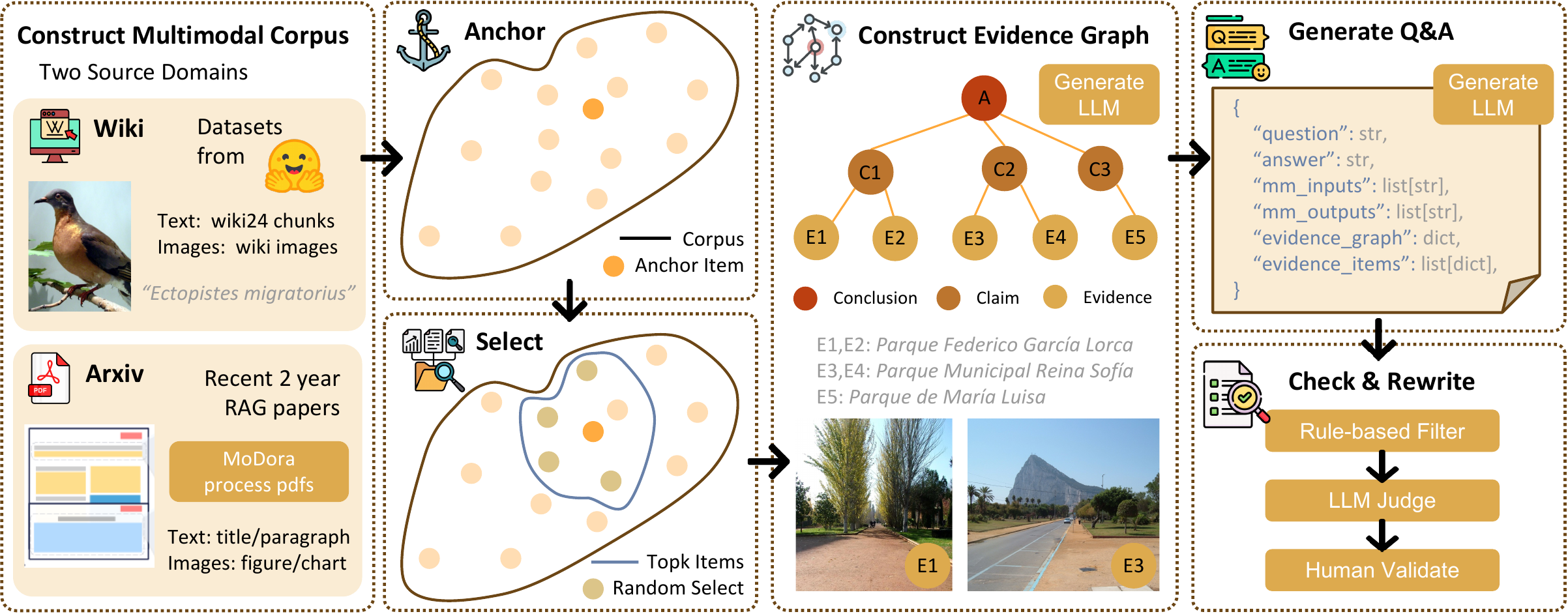}  
    \caption{Pipeline for Constructing HiEviDR-Bench. We first build a multimodal corpus from two source domains, then select anchor-centered candidate evidence items to construct an evidence graph with an LLM. Based on the graph, we generate question--answer pairs and apply rule-based filtering, Judge-Model checking, and human validating to ensure structural validity and data quality.}
  \label{fig:benchmark}
\end{figure*}

\textbf{Citation Score.}
Citation Score evaluates whether the recalled key evidence is correctly cited and grounded in the report.
This stage is activated only when the Citation Gate is satisfied.
Let $\mathcal{N}_{\text{e}}$ denote the set of evidence nodes in the evidence graph $\mathcal{G}$, and let $\mathcal{K}_{e}$ denote the set of citation-quality dimensions, including Accuracy (Acc.), Consistency (Cons.), Structure (Struc.), and Integration (Integ.).

For each evidence node $n \in \mathcal{N}_{\text{e}}$, the binary variable $w_n \in \{0,1\}$ indicates whether the corresponding evidence item is cited in the report.
If $w_n=1$, we pair the evidence item of node $n$ with its citation context in the report and ask the Judge-Model to assign a score $r_{n,k}\in[1,5]$ for each dimension $k \in \mathcal{K}_{e}$.
The final Citation Score is
\begin{equation}
S_{\text{citation}}= 20 \cdot \mathrm{Gate}(\text{citation}) \cdot
\frac{1}{5|\mathcal{N}_{\text{e}}||\mathcal{K}_{e}|}
\sum_{n \in \mathcal{N}_{\text{e}}}w_n\cdot \left(
\sum_{k \in \mathcal{K}_{e}}
r_{n,k}\right),
\label{eq:citation}
\end{equation}
where $|\mathcal{N}_{e}|$ denotes the number of nodes $\mathcal{N}_{e}$ in evidence graph $\mathcal{G}$ and $|\mathcal{K}_{e}|$ denotes the number of dimensions in $\mathcal{K}_{e}$.

\textbf{Claim Score.}
Claim Score evaluates whether the intermediate claims in the report are well-grounded by the cited evidence and faithfully reflect the evidence graph.
This stage is activated only when the Claim Gate is satisfied.
Let $\mathcal{N}_{\text{c}}$ denote the set of claim nodes in the evidence graph $\mathcal{G}$, and let $\mathcal{K}_{c}$ denote the set of claim-quality dimensions, including corresponding to Relevance (Rel.), Contextualization (Ctx.), and Clarity (Clr.).

For each claim node $n \in \mathcal{N}_{\text{c}}$, the binary variable $w_n \in \{0,1\}$ indicates whether the corresponding claim is instantiated and supported in the report.
If $w_n=1$, the associated claim span is scored by the Judge-Model on each dimension $k \in \mathcal{K}_{c}$, yielding $r_{n,k}\in[1,5]$.
The final Claim Score is
\begin{equation}
S_{\text{claim}}
=
20 \cdot \mathrm{Gate}(\text{claim}) \cdot
\frac{1}{5|\mathcal{N}_{\text{c}}||\mathcal{K}_{c}|}
\sum_{n \in \mathcal{N}_{\text{c}}} w_n\cdot \left(
\sum_{k \in \mathcal{K}_{c}}
 r_{n,k}\right),
\label{eq:claim}
\end{equation}
where  $|\mathcal{N}_{c}|$ denotes the number of claim nodes $\mathcal{N}_{c}$ in evidence graph $\mathcal{G}$ and $|\mathcal{K}_{c}|$ denotes the number of dimensions in $\mathcal{K}_{c}$.

\textbf{Answer Score.}
Answer Score evaluates the final report conclusion, which is activated only when the Answer Gate is satisfied---that is, all required claims are supported, which can be interpreted as the full activation of $\mathcal{G}$ up to the conclusion node $n_{\text{conclusion}}$.
We extract the answer conclusion from the report and ask the Judge-Model to assess it along multiple answer-quality dimensions.
Let $\mathcal{K}_{a}$ denote the set of answer-quality dimensions, including Accuracy (Acc.), Coherence (Coh.), Integration (Integ.), Completeness (Comp.), Depth, and Conciseness (Conc.).

Let $r_k \in [1,5]$ be the Judge-Model score for dimension $k \in \mathcal{K}_{a}$.
The final Answer Score is
\begin{equation}
S_{\text{answer}}
=
20 \cdot \mathrm{Gate}(\text{answer}) \cdot
\frac{1}{5|\mathcal{K}_{a}|}\sum_{k \in \mathcal{K}_{a}} r_k,
\label{eq:answer}
\end{equation}
where $|\mathcal{K}_{a}|$ denotes the number of dimensions in $\mathcal{K}_{a}$.

Overall, the proposed evaluation framework transforms report assessment from the single-step quality check of Eq.~\ref{eq:existing_eval} into a structured, traceability-oriented measurement grounded in the evidence graph $\mathcal{G}$.
By requiring that every scoring dimension is anchored to specific nodes and edges in the graph, our benchmark can distinguish between methods that build conclusions through faithful evidence aggregation and those that only generate superficially convincing responses.

\begin{table*}[t]
\centering
\caption{Dataset statistics of the HiEviDR-Bench. \#Avg. Nodes indicates the averaged node number in evidence graph.}
\label{tab:benchmark_statistics}
\setlength{\tabcolsep}{10.2pt}
\begin{tabular}{llcccccc}
\toprule
\multirow{2}{*}{\textbf{Domain}} & \multirow{2}{*}{\textbf{Modality}} 
& \multicolumn{2}{c}{\textbf{Easy}} 
& \multicolumn{2}{c}{\textbf{Medium}} 
& \multicolumn{2}{c}{\textbf{Hard}} \\
\cmidrule(lr){3-4} \cmidrule(lr){5-6} \cmidrule(lr){7-8}
& & \textbf{\#Samples} & \textbf{\#Avg. Nodes} 
  & \textbf{\#Samples} & \textbf{\#Avg. Nodes} 
  & \textbf{\#Samples} & \textbf{\#Avg. Nodes} \\
\midrule
\multirow{2}{*}{Wiki}
& Text       & 155 & 7.76 & 195 & 10.79 & 150 & 16.84 \\
& Multimodal & 225 & 5.52 & 135 & 7.00 & 140 & 8.75 \\
\midrule
\multirow{2}{*}{Arxiv}
& Text       & 204 & 7.18 & 158 & 10.15 & 138 & 14.17 \\
& Multimodal & 222 & 6.41 & 140 & 8.93 & 138 & 13.07 \\
\bottomrule
\end{tabular}
\end{table*}

\subsection{Data Curation}
\label{sec:data_curation}

The core of HiEviDR-Bench is the evidence graph $\mathcal{G}$, which provides the structured supervision signal for the traceability-oriented evaluation in Sec.~\ref{sec:eval_dim}.
As illustrated in Figure~\ref{fig:benchmark}, we first describe the design of $\mathcal{G}$, then present the four-stage pipeline that produces it: corpus construction, evidence collection, graph construction, and question--answer generation. Although the pipeline is largely LLM-assisted, all constructed evidence graphs and question–answer pairs are further verified by human annotators before inclusion in the final benchmark.

\textbf{Evidence Graph Design.}
As defined in Eq.~\ref{eq:graph}, the evidence graph $\mathcal{G} = (\mathcal{N}, \mathcal{E})$ is a directed acyclic graph whose node set $\mathcal{N} = \mathcal{N}_{\text{e}} \cup \mathcal{N}_{\text{c}} \cup \{n_{\text{conclusion}}\}$ organizes heterogeneous information into a hierarchical reasoning structure and the edge set $\mathcal{E} \subseteq \mathcal{N} \times \mathcal{N}$ captures directed support relations.
Evidence nodes ($\mathcal{N}_{\text{e}}$) correspond to atomic textual or visual evidence items from the corpus.
Claim nodes ($\mathcal{N}_{\text{c}}$) represent intermediate reasoning statements grounded in one or more evidence or lower-level claim nodes.
The conclusion node ($n_{\text{conclusion}}$) denotes the final answer supported by the full graph, connected to one or more claim nodes through directed edges representing the top-level reasoning step.
Each directed edge $(n_i, n_j) \in\mathcal{E}$ uniformly indicates that $n_i$  provides direct evidential or logical support for $n_j$.

This design serves two purposes: on the evaluation side, each scoring dimension in Sec.~\ref{sec:eval_dim} is anchored to specific nodes and edges, enabling fine-grained diagnosis of where the reasoning chain breaks down; on the data construction side, the graph acts as a generative scaffold, ensuring that every benchmark instance requires multi-step evidence aggregation rather than single-hop retrieval.

In practice, the graph size varies with task difficulty: 
Easy instances contain graphs with an average of 5--8 nodes arranged in 2 reasoning layers, while Hard instances contain 9--17 nodes spanning up to 4 layers.
This progression in graph complexity directly governs the difficulty of evidence aggregation required to answer the corresponding question (evidence to low-level claim to high-level claim to conclusion).

\textbf{Multimodal Corpus and Evidence Collection.}
We build a large multimodal corpus from two complementary domains: Wikipedia, which provides broad open-domain knowledge with text chunks and associated images~\cite{wikiimage_dataset, ultrarag_benchmark}, and arXiv, which provides recent RAG-related papers parsed by MoDora~\cite{xu2026modora} into textual and visual content for domain-specific evaluation. This dual-domain setup allows HiEviDR-Bench to evaluate evidence aggregation in both open-domain and specialized academic settings.

For each instance, we select one corpus item as an anchor, retrieve the top-$k$ relevant items with Qwen3-VL-Embedding-2B~\cite{qwen3vlembedding}, and further sample from its neighborhood for diversity, yielding a candidate set of 10--25 textual and visual evidence items.

\textbf{Graph Construction Pipeline.}
Given the candidate evidence set, we construct the evidence graph $\mathcal{G}$ with an LLM under a structured schema that specifies the three node types and directed support relations. The model is instructed to identify relations among evidence items, derive intermediate claims supported by subsets of evidence, and synthesize them into a final conclusion.

The resulting graph specification includes evidence nodes, claim nodes, a conclusion node, and their support edges. We then apply rule-based validation to ensure that $\mathcal{G}$ is a connected directed acyclic graph with a single conclusion root and that each evidence node contributes to at least one claim. Invalid graphs are discarded and regenerated.

\textbf{Question--Answer Generation and Quality Control.}
Given each validated graph $\mathcal{G}$ and its evidence items, we prompt an LLM to generate a question--answer pair that requires multi-node evidence aggregation. Each instance contains six fields: \textit{question}, \textit{answer}, \textit{mm\_inputs}, \textit{mm\_outputs}, \textit{evidence\_graph}, and \textit{evidence\_items}, preserving both multimodal content and structured supervision.

We perform quality control in three stages. Starting from 30,000 automatically constructed instances, we first apply rule-based checks to remove structurally invalid samples, retaining 12,853 candidates. We then employ a judge model to evaluate logical soundness and question–answer alignment, further reducing the dataset to 3,407 high-quality instances. Finally, we conduct manual validation to verify the correctness and reliability of the remaining samples, selected 500 instances from each domain-modality paired, resulting in a final benchmark of 2,000 instances.

\textbf{Data Statistics.}
Table~\ref{tab:corpus_statistics} reports the corpus-level statistics, and Table~\ref{tab:benchmark_statistics} presents the benchmark-level statistics.
To support evaluation of both text-only and multimodal deep research methods, the benchmark is divided into two modality settings.
Samples are further categorized into three difficulty levels based on the number and density of nodes in the evidence graph $\mathcal{G}$: the average graph size increases consistently from Easy to Hard, indicating that the difficulty partition also corresponds to increasing evidence aggregation complexity.
For each domain, modality, and difficulty level, we report the number of samples and the average number of graph nodes. 
Further details on the data curation, including human validation and comparisons with existing benchmarks, are provided in Appendix~\ref{sec:human_validation} and Appendix~\ref{sec:comparison}, respectively.

\begin{table*}[t]
\centering
\small
\setlength{\tabcolsep}{4.5pt}
\caption{Overall results of HiEviDR-Bench. The best scores in each column are highlighted, and the secondary scores are underlined. Report, Trace., Citation, Claim, and Answer denote the five evaluation dimensions.}
\begin{tabular}{lccccccccccccc}
\toprule
\multirow{3}{*}{\textbf{Method}} 
& \multicolumn{6}{c}{\textbf{Wiki}} 
& \multicolumn{6}{c}{\textbf{Arxiv}} 
& \multirow{3}{*}{\textbf{Avg.}} \\
\cmidrule(lr){2-7} \cmidrule(lr){8-13}
& Report & Trace. & Citation
& Claim & Answer & Overall
& Report & Trace. & Citation
& Claim & Answer & Overall
& \\
\midrule

\rowcolor{gray!12}
\multicolumn{14}{c}{\textbf{MLLM with RAG}} \\
\midrule
Qwen3.5-4B              & 18.58 & 9.01 & 4.64 & 2.66 & 1.15 & 36.04 & 19.62 & 6.10 & 3.22 & 2.40 & 0.63 & 31.97 & 34.00 \\
MiMo-VL-7B-RL           & 18.36 & 8.76 & 6.13 & 3.23 & 1.37 & 37.84 & 19.05 & 5.89 & 3.30 & 2.46 & 0.64 & 31.33 & 34.58 \\
MiniCPM-V-4.5           & 17.63 & 8.72 & 4.01 & 2.50 & 0.99 & 33.84  & 18.22 & 6.55 & 1.75 & 1.31 & 0.27 & 28.06 & 30.95 \\
InternVL3.5-8B          & 18.23 & 9.10 & 7.07 & 3.79 & 1.72 & 39.90 & 18.66 & \underline{6.64} & 3.86 & 2.75 & 0.60 & 32.50 & 36.20 \\
GLM-4.6V-Flash          & 17.52 & 9.07 & 3.75 & 2.44 & 1.15 & 33.93 & 18.06 & 6.41 & 3.22 & 2.42 & 0.64 & 30.73 & 32.33 \\
Gemma4-E4B-it           & 17.73 & 8.67 & 6.90 & 4.09 & 1.78 & 39.16 & 18.77 & 5.68 & 4.19 & 3.03 & 0.84 & 32.49 & 35.83 \\
Qwen3-VL-8B-Instruct    & 18.33 & \underline{9.13} & 6.25 & 3.58 & 1.63 & 38.90 & 19.04 & 6.24 & 3.62 & 2.70 & 0.59 & 32.18 & 35.54 \\
Qwen3.5-9B              & 18.50 & 8.69 & 4.71 & 2.88 & 1.31 & 36.07 & 19.29 & 6.31 & 2.38 & 1.76 & 0.46 & 30.19 & 33.13 \\
Qwen3.5-35B-A3B         & 18.56 & 8.62 & 5.62 & 3.42 & 1.51 & 37.72 & 19.36 & 5.46 & 3.22 & 2.37 & 0.43 & 30.82 & 34.27 \\
Qwen3.5-27B             & 18.58 & 8.58 & 6.75 & 3.83 & 1.63 & 39.35 & 19.28 & 5.66 & 4.11 & 3.03 & 0.76 & 32.83 & 36.09 \\
Gemma4-31B-it           & 18.28 & 8.91 & 8.02 & \underline{4.49} & 2.03 & 41.71 & 19.26 & 5.71 & 4.45 & 3.45 & \underline{0.96} & \underline{33.82} & \underline{37.76} \\
GPT-5-mini              & 18.81 & 8.60 & 6.76 & 4.22 & 1.79 & 40.18 & \cellcolor{blue!15}\textbf{19.80} &	5.44 & 3.62 & \cellcolor{blue!15}\textbf{3.68} & 0.75 &	32.80 &	36.74 \\ 
Gemini-3.1-Flash        & 18.32 & 8.77 & 7.77 & 4.02 & 1.70 & 40.59 & 19.16 & 5.65 & 4.44 & 3.27 & 0.69 & 33.21 & 36.90 \\
GPT-5                   & 18.42 & 8.93 & \cellcolor{blue!15}\textbf{8.41} & 4.26 & 1.86 & \underline{41.89} & 19.20 & 5.08 & \cellcolor{blue!15}\textbf{4.72} & \underline{3.51} & 0.83 & 33.34 & 37.62 \\
GPT-5.6-sol             & \cellcolor{blue!15}\textbf{18.88} & 8.92 & \underline{8.21} & \cellcolor{blue!15}\textbf{4.53} & \underline{2.14} & \cellcolor{blue!15}\textbf{42.67} & \underline{19.69} & 5.31 & \underline{4.55} & \underline{3.51} & \cellcolor{blue!15}\textbf{0.97} & \cellcolor{blue!15}\textbf{34.03} & \cellcolor{blue!15}\textbf{38.35} \\
Grok-4.5                & \underline{18.66} & \cellcolor{blue!15}\textbf{9.43} & 6.61 & 3.90 & \cellcolor{blue!15}\textbf{2.22} & 40.82 & 19.43 & \cellcolor{blue!15}\textbf{6.79} & 3.33 & 2.54 & 0.71 & 32.79 & 36.82 \\
\midrule

\rowcolor{gray!12}
\multicolumn{14}{c}{\textbf{MLLM with Deep Research}} \\
\midrule
Qwen3.5-4B              & 19.07 & 9.11 & 6.70 & 3.87 & 1.61 & 40.35  & 19.61 & 6.83 & 3.70 & 3.21 & 0.92 & 34.26 & 37.31 \\
MiMo-VL-7B-RL           & 18.00 & 8.97 & 5.36 & 2.43 & 0.97 & 35.73 & 18.53 & 6.75 & 3.02 & 2.35 & 0.74 & 31.33 & 33.53 \\
MiniCPM-V-4.5           & 18.56 & 9.50 & 3.86 & 2.37 & 1.10 & 35.37  & 18.97 & \cellcolor{blue!15}\textbf{7.26} & 3.06 & 2.67 & 0.89 & 32.79 & 34.08 \\
InternVL3.5-8B          & 18.23 & 9.64 & 5.57 & 3.37 & 1.31 & 38.07 & 18.56 & 7.00 & 4.01 & 2.88 & 0.80 & 33.23 & 35.65 \\
GLM-4.6V-Flash          & 18.28 & 9.45 & 4.42 & 2.87 & 1.21 & 36.21 & 18.86 & 7.24 & 3.59 & 2.92 & 0.90 & 33.50 & 34.85 \\
Gemma4-E4B-it           & 18.84 & 9.16 & 6.33 & 4.11 & 1.82 & 40.25 & 19.63 & 6.35 & 3.40 & 3.08 & 0.83 & 33.27 & 36.76 \\
Qwen3-VL-8B-Instruct    & 19.08 & 9.64 & 5.92 & 3.16 & 1.34 & 39.12 & 19.65 & \underline{7.18} & 4.15 & 3.22 & 0.91 & 35.10 & 37.11 \\
Qwen3.5-9B              & 18.90 & 9.35 & 6.08 & 3.91 & 1.68 & 39.88 & 19.66 & 6.80 & 4.04 & 3.52 & 1.11 & 35.11 & 37.50 \\
Qwen3.5-35B-A3B         & 19.22 & 9.43 & 6.89 & 4.42 & 1.98 & 41.94 & \cellcolor{blue!15}\textbf{19.78} & 6.47 & 3.99 & 3.40 & 0.99 & 34.61 & 38.28 \\
Qwen3.5-27B             & \cellcolor{blue!15}\textbf{19.50} & 9.16 & 6.97 & 4.68 & 2.13 & 42.44 & 19.71 & 6.46 & 3.69 & 3.67 & 1.13 & 34.65 & 38.54 \\
Gemma4-31B-it           & 19.04 & 9.41 & 7.93 & 5.01 & 2.31 & 43.71 & 19.58 & 6.31 & 4.11 & \underline{3.74} & 1.14 & 34.87 & 39.29 \\
GPT-5-mini              & 18.87 & 8.89 & 8.39 & 4.19 & 1.90 & 42.25 & 19.56 & 5.25 & \cellcolor{blue!15}\textbf{4.70} & 3.39 & 1.10 &	34.00 &	38.13 \\ 
Gemini-3.1-Flash        & 19.10 & \underline{9.70} & 7.76 & 4.68 & 1.86 & 43.10 & 19.49 & 6.94 & 4.02 & 3.59 & 1.05 & 35.09 & 39.10 \\ 
GPT-5                   & 18.95 & 9.32 & \underline{8.66} & 4.25 & 2.13 & 43.31 & 19.45 & 6.48 & \underline{4.57} & 3.29 & 1.08 & 34.87 & 39.09 \\ 
GPT-5.6-sol             & \underline{19.33} & 9.63 & 8.40 & \underline{5.18} & \underline{2.60} & \underline{45.13} & \underline{19.73} & 6.32 & 3.90 & \cellcolor{blue!15}\textbf{3.86} & \cellcolor{blue!15}\textbf{1.33} & \underline{35.13} & \underline{40.13} \\
Grok-4.5                & 19.00 & \cellcolor{blue!15}\textbf{9.87} & \cellcolor{blue!15}\textbf{9.21} & \cellcolor{blue!15}\textbf{5.40} & \cellcolor{blue!15}\textbf{2.94} & \cellcolor{blue!15}\textbf{46.41} & 18.90 & 6.76 & 4.52 & 3.62 & \underline{1.17} & \cellcolor{blue!15}\textbf{35.95} & \cellcolor{blue!15}\textbf{41.18} \\
\bottomrule
\end{tabular}
\label{tab:overall}
\end{table*}

\section{Results and Analysis}

In this section, we first present the overall evaluation results on HiEviDR-Bench, followed by analyses of evidence traceability and evidence aggregation capabilities.
Additional analyses across different domains and modalities, together with qualitative case studies, are provided in Appendix~\ref{sec:domain} and Appendix~\ref{sec:case}, respectively.

\subsection{Experimental Setup}

\textbf{Evaluated Models.}
We evaluate representative state-of-the-art MLLMs under two report-generation settings: (i) \emph{MLLM with RAG} and (ii) \emph{MLLM with Deep Research}. 
The evaluated models include the Qwen-VL family~\cite{qwen3technicalreport,qwen3_5}, Gemma4 family~\cite{gemmateam2026gemma4}, the GPT-5 family~\cite{singh2025openai}, the Gemini-3.1-Flash~\cite{team2023gemini}, Grok-4.5~\cite{xai2026grok45}, GLM-4.6v-Flash~\cite{vteam2025glm45vglm41vthinkingversatilemultimodal}, InternVL3.5-8B~\cite{wang2025internvl3_5}, MiMo-VL-7B-RL~\cite{coreteam2025mimovltechnicalreport}, and MiniCPM-V-4.5~\cite{yao2024minicpm}.

\textbf{Implementation Details.}
For corpus construction, we use Qwen3-VL-Embedding-2B~\cite{qwen3vlembedding} as the embedding model. 
Under RAG, we retrieve the top-15 results as supplementary evidence. 
Under Deep Research, each model is allowed up to three rounds of iterative question refinement; for each sub-problem, we retrieve the top-20 results, then keep the top-25 text chunks and top-5 images after re-ranking. 
For all LLM-based scoring steps, we use Qwen3-VL-235B-A22B-Instruct as the judge model with temperature 0.2.
More evaluation details are shown in Appendix~\ref{sec:app_evaluation}.

\begin{figure*}[t!]
  \centering
  \includegraphics[width=1.0\textwidth]{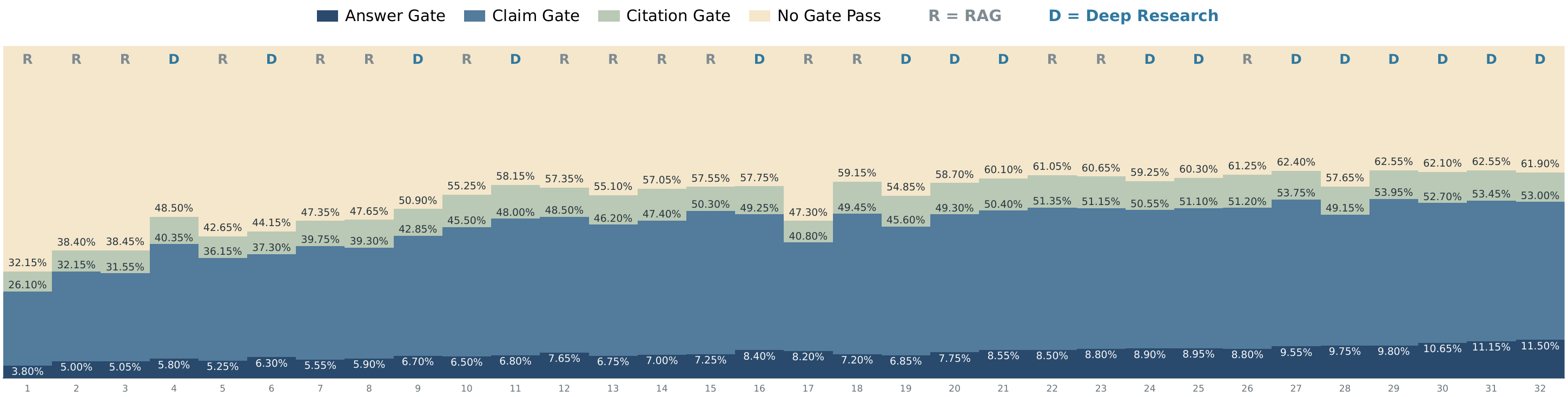}  
\caption{Stage-wise gate pass rates under the progressive evaluation pipeline.
Each stacked bar shows the proportion of samples that pass the Answer, Claim, and Citation gates, or fail at earlier stages (No Gate Pass).
Models are ordered from left to right by their overall benchmark scores, with the index-to-model mapping provided in Appendix Table~\ref{tab:rate_index_mapping}.}
  \label{fig:gate}
\end{figure*}
\begin{figure}[t!]
  \centering
  \includegraphics[width=\linewidth]{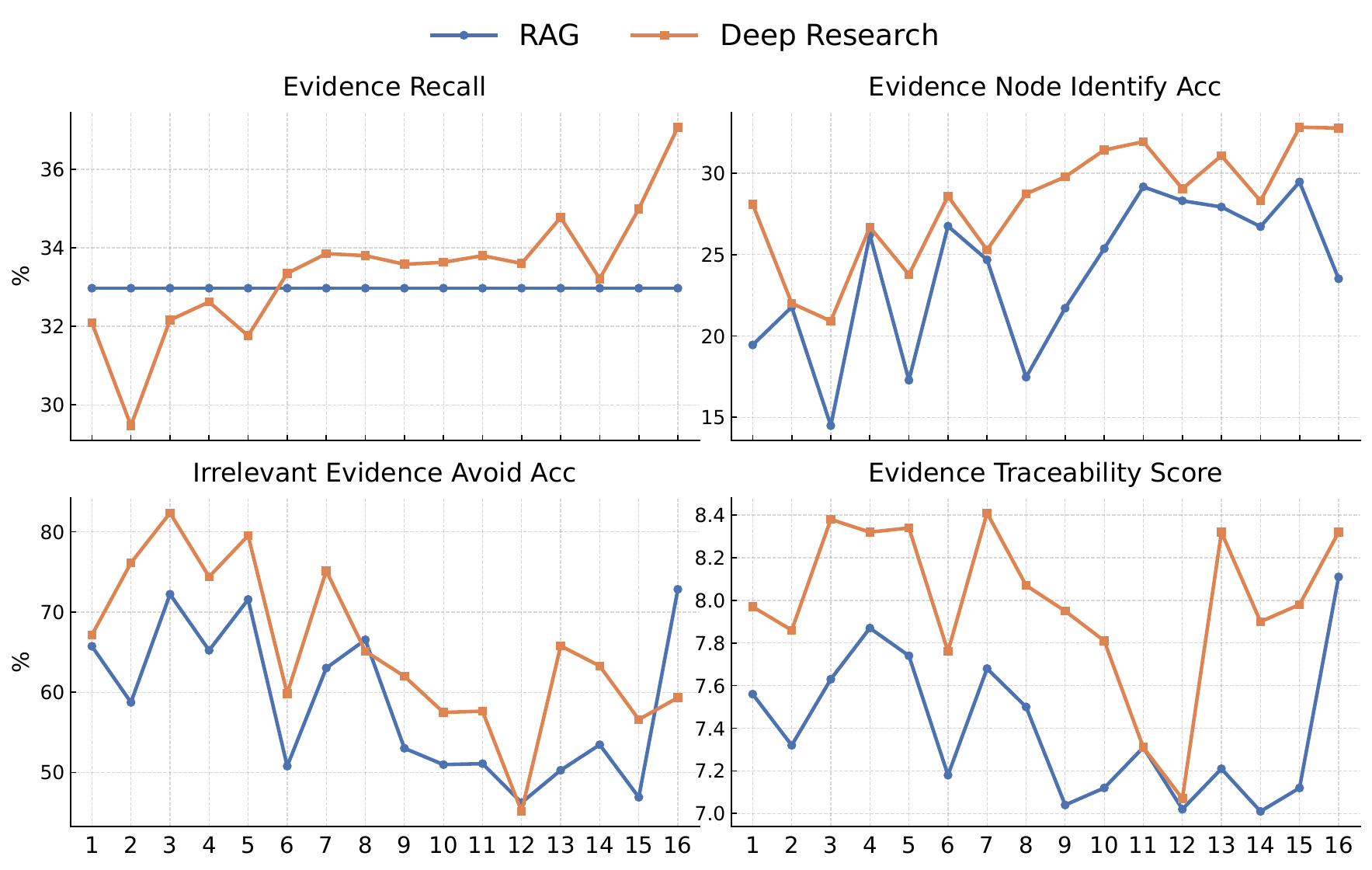}  
\caption{Comparison of RAG and Deep Research on four evidence traceability-related metrics. Models are indexed from 1 to 16 for readability; the index-to-model mapping is provided in Appendix Table~\ref{tab:model_index_mapping}.}
  \label{fig:evidence}
\end{figure}

\subsection{Main Results and Findings}

Table~\ref{tab:overall} reports the overall results on HiEviDR-Bench under both \textit{MLLM with RAG} and \textit{MLLM with Deep Research} settings. Overall, Deep Research generally achieves stronger performance than vanilla RAG, suggesting that explicit evidence integration is beneficial for grounded multimodal report generation. However, the results also reveal a clear mismatch between report quality and evidence-grounded reasoning. While many models obtain relatively strong Report scores, their performance drops substantially on Citation, Claim, and Answer, indicating that producing a fluent and well-structured report is considerably easier than faithfully organizing evidence into supported intermediate claims and grounded conclusions.

This gap is consistent across model families and highlights the main value of HiEviDR-Bench. A model may generate a report that appears coherent and informative, yet still fail to correctly attribute evidence, construct supported claims, or reach a justified final answer. Stronger models also tend to distinguish themselves more clearly on these later evidence-grounded dimensions than on report quality alone, suggesting that the key challenge of multimodal deep research lies less in surface-level fluency and more in hierarchical evidence aggregation.

\subsection{Evidence Aggregation Analysis}

Figure~\ref{fig:gate} presents the stage distribution under the progressive gating pipeline. Overall, the stage distribution is broadly consistent with the final ranking: stronger models tend to perform better at later gates, especially those related to claim composition and answer generation.

More importantly, the most dominant failure patterns across models are concentrated in No Gate Pass and Claim Gate. This suggests that current systems face two main bottlenecks: identifying truly critical evidence from a large candidate pool, and organizing that evidence into sufficiently supported intermediate claims. By comparison, the proportion at Answer Gate is relatively smaller, indicating that once evidence has been properly organized, generating the final answer is not necessarily the hardest step.

In addition, for several model pairs, Deep Research improves intermediate gate performance even when the final overall gain is limited. This suggests that its benefits first appear in the evidence aggregation process itself, although such improvements are not always fully reflected in final report quality.

\subsection{Evidence Traceability Analysis}

Figure~\ref{fig:evidence} further compares RAG and Deep Research on Evidence Recall, Evidence Node Identify Acc, Irrelevant Evidence Avoid Acc, and Evidence Traceability Score. The results show that the advantage of Deep Research does not mainly come from recalling more relevant evidence. In fact, it does not consistently outperform RAG on Evidence Recall or Evidence Node Identify Acc, and for some backbones it is even weaker. This suggests that simply expanding the search process or exploring a larger evidence space does not automatically improve the quality of evidence use.

Its clearer advantage instead appears in Evidence Node Identify Acc and Irrelevant Evidence Avoid Acc, together with higher or comparable Evidence Traceability Score. This indicates that the gains of Deep Research come more from suppressing distracting evidence and preserving more meaningful evidence-to-report links than from improving retrieval coverage alone. Put differently, Deep Research is more helpful in controlling and organizing evidence than in merely finding more of it.

Taken together, these findings suggest that evidence traceability depends not only on whether relevant evidence is retrieved, but also on whether irrelevant evidence is filtered out and whether the retained evidence is meaningfully connected to the final report. This further supports the central conclusion of HiEviDR-Bench: the key challenge of multimodal deep research lies less in evidence retrieval itself and more in hierarchical evidence aggregation.
\section{Conclusion}

We present \textbf{HiEviDR-Bench}, a traceability-oriented benchmark for evaluating hierarchical evidence aggregation in multimodal deep research. It enables fine-grained diagnosis of how models retrieve, organize, and compose heterogeneous evidence into grounded reports. Experiments on representative MLLMs show that current systems still struggle with evidence traceability, claim-level reasoning, and faithful answer generation.

\bibliographystyle{ACM-Reference-Format}
\bibliography{main}

\clearpage
\appendix
\section{Appendix}

\subsection{Human Validation Details of Datasets}
\label{sec:human_validation}

Although QA pairs and evidence graphs are constructed with LLM assistance, they are not directly accepted from LLM outputs. 
Our pipeline applies strict rule-based filtering and consistency checks to remove noisy, incomplete, or weakly grounded samples; from 30,000 randomly selected anchor evidence items, only 3,407 pass all filters, resulting in an acceptance rate of \textbf{11.36\%}. 

\begin{figure}[h]
  \centering
  \includegraphics[width=\linewidth]{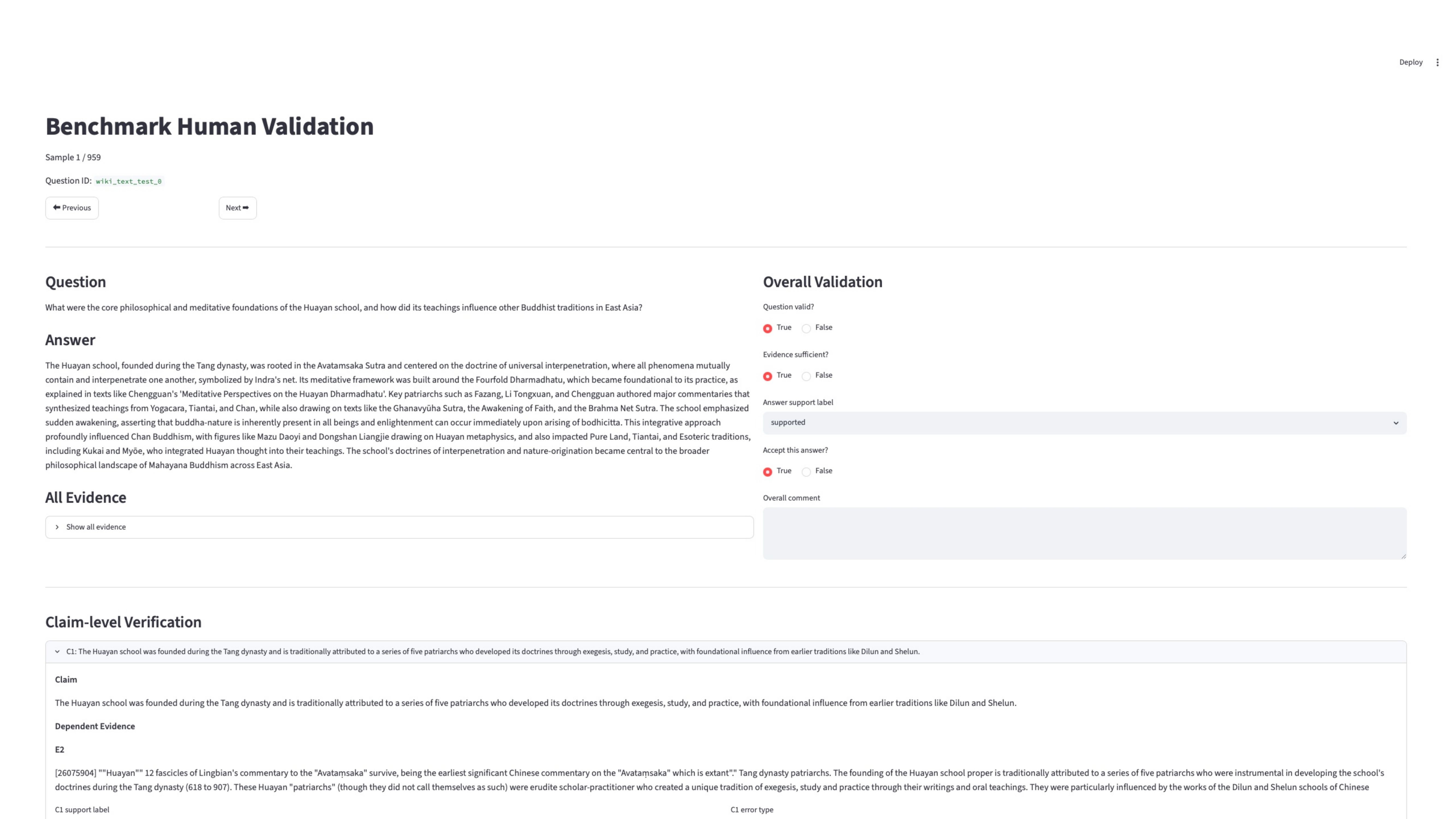}  
    \caption{Human validation interface.}
  \label{fig:annotation}
\end{figure}

As shown in Figure~\ref{fig:annotation}, we further conduct human validation on the filtered samples, and finally selected 2000 samples evenly from four subdomains.
Two annotators verify question validity, answer support, claim grounding, evidence-to-claim edges, claim-to-conclusion edges, and graph-level acceptance. 
Table~\ref{tab:human_validation_summary} shows high adjudicated accuracy, raw agreement, and Cohen's $\kappa$, indicating that the filtered evidence graphs are reliable; the few remaining errors mainly involve ambiguous questions, incomplete evidence, or mildly over-generalized claims, while incorrect graph links are rare. 

\begin{table}[h]
\centering
\caption{Human validation data statics.}
\label{tab:human_validation_summary}
\small
\setlength{\tabcolsep}{10pt}
\begin{tabular}{lccc}
\toprule
Target & Acc. & Agree. & $\kappa$ \\
\midrule
Question & 98.5 & 99.0 & 0.86 \\
Answer & 99.3 & 99.5 & 0.91 \\
Claim & 99.8 & 99.6 & 0.94 \\
E2C & 99.8 & 99.7 & 0.95 \\
C2A & 100.0 & 100.0 & 1.00 \\
Graph & 97.3 & 98.0 & 0.83 \\
\bottomrule
\end{tabular}
\end{table}

\subsection{Benchmark novelty and comparison with prior work}
\label{sec:comparison}

Table~\ref{tab:benchmark_comparison} shows that prior benchmarks cover browsing, long-form reporting, citation evaluation, interaction, or multimodal reasoning, while our core novelty is the explicit evidence-to-claim-to-conclusion graph for process-level grounding evaluation.

\begin{table*}[h]
\centering
\caption{Comparison with related benchmarks.}
\label{tab:benchmark_comparison}
\small
\setlength{\tabcolsep}{11pt}
\begin{tabular}{lccccccc}
\toprule
Benchmark & Multimodal & Corpus & Long Response & Citation & Claim & Procedure & Evidence Graph \\
\midrule
ViDoSeek~\cite{wang2025vidorag} & \checkmark & \checkmark & -- & -- & -- & -- & -- \\
ALCE~\cite{gao2023enabling} & -- & \checkmark & \checkmark & \checkmark & \checkmark & -- & -- \\
LongBench-Cite~\cite{zhang2025longcite} & -- & \checkmark & -- & \checkmark & \checkmark & -- & -- \\
IDRBench~\cite{feng2026idrbench} & -- & -- & \checkmark & -- & -- & -- & \checkmark \\
DeepResearch Bench~\cite{du2025deepresearch} & -- & -- & \checkmark & \checkmark & \checkmark & -- & \checkmark \\
BrowseComp~\cite{wei2025browsecomp} & -- & -- & -- & -- & -- & -- & \checkmark \\
BrowseComp-Plus~\cite{chen2025browsecomp} & -- & \checkmark & -- & \checkmark & -- & -- & \checkmark \\
MMDR-Bench~\cite{huang2026mmdeepresearch} & \checkmark & -- & \checkmark & \checkmark & \checkmark & -- & \checkmark \\
\midrule
Ours & \checkmark & \checkmark & \checkmark & \checkmark & \checkmark & \checkmark & \checkmark \\
\bottomrule
\end{tabular}
\end{table*}

\subsection{Additional Pipeline Details}

To facilitate reproducibility, we provide additional details of the two report generation settings used in our experiments, namely \textit{MLLM with RAG} and \textit{MLLM with Deep Research}. Although both settings ultimately produce an HTML-formatted multimodal report, they differ substantially in how evidence is gathered, organized, and incorporated before the final rendering stage.

\begin{figure}[t]
  \centering
  \includegraphics[width=1.0\linewidth]{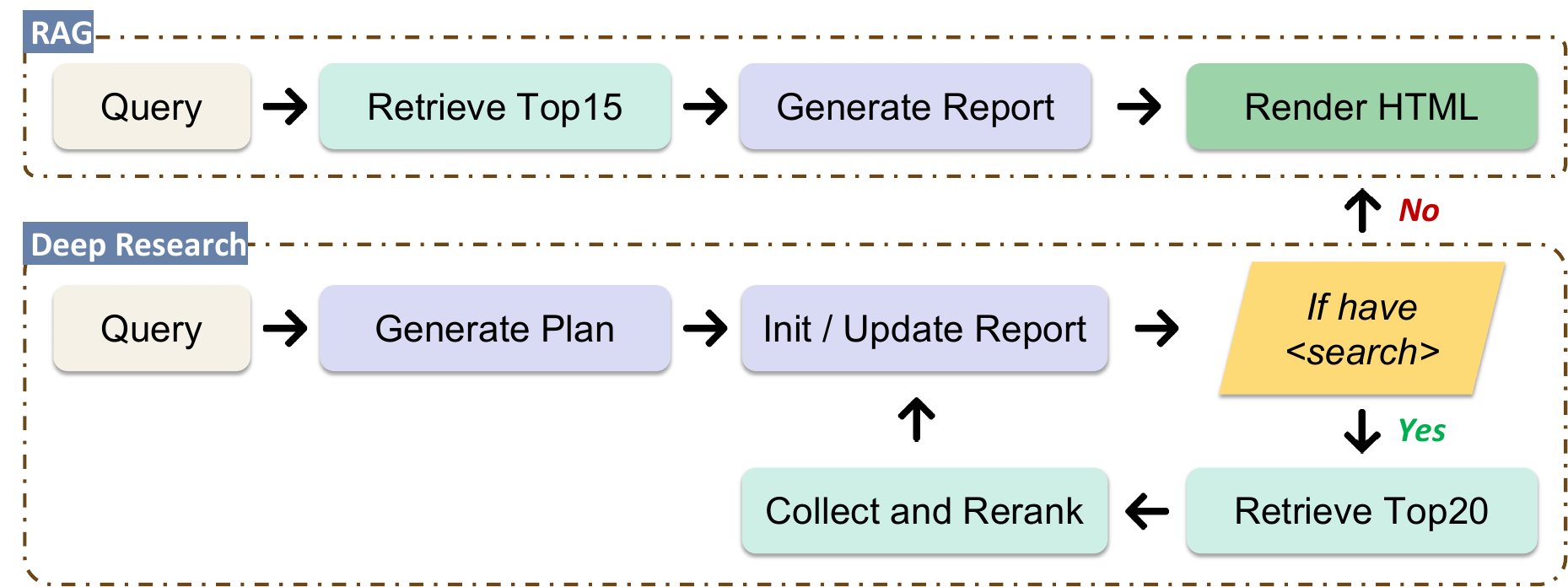}  
\caption{Overview of the RAG and Deep Research pipelines used for multimodal report generation.}
  \label{fig:pipeline}
\end{figure}

Figure~\ref{fig:pipeline} presents an overview of the two pipelines. In the RAG setting, the model receives the user query together with a fixed set of retrieved multimodal evidence, and then directly generates the final report in a single pass. This design follows a standard retrieval-augmented generation paradigm, where the retrieved context is fixed before generation and report writing is performed immediately afterward. In contrast, the Deep Research pipeline introduces an iterative evidence construction process before final rendering. Starting from the initial query, the model first generates a research plan, then performs retrieval according to intermediate search needs, collects and reranks the returned evidence, and continuously updates the report. Compared with RAG, this procedure enables the model to progressively expand, refine, and reorganize its evidence basis, rather than relying solely on the initially retrieved context.

\begin{figure*}[t!]
  \centering
  \includegraphics[height=0.81\textheight]{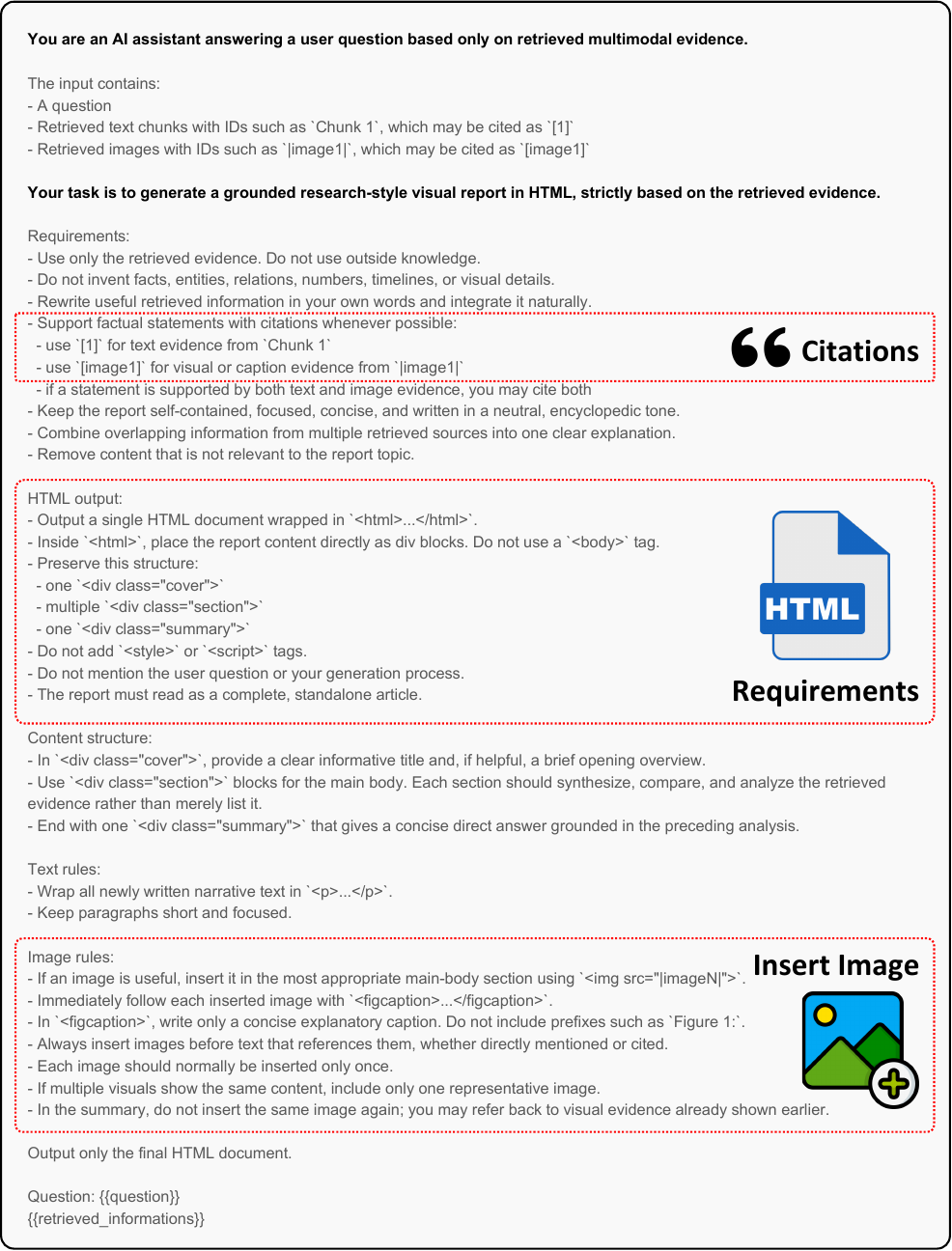}  
\caption{Prompt template used in the RAG setting.}
  \label{fig:rag_prompt}
\end{figure*}

Figure~\ref{fig:rag_prompt} further shows the prompt template used in the RAG setting. We use this prompt to illustrate how the multimodal report is constructed and rendered in our framework. Specifically, the prompt requires the model to generate a grounded research-style report in HTML format based only on the retrieved text chunks and images, while prohibiting the use of outside knowledge. It also explicitly specifies citation rules for textual and visual evidence, so that the model can properly attribute evidence from text chunks and image captions in the generated report. Moreover, retrieved visual evidence is controlled through image placeholders such as \texttt{|imagek|}, which allow the model to select and insert specific images into the HTML output. The resulting HTML document is then rendered into the final multimodal report, where textual analysis, cited evidence, and inserted images are presented in a unified format.

Overall, the essential difference between the two settings lies in whether evidence organization is performed only once or iteratively. RAG can be viewed as direct grounded generation over a fixed retrieved context, while Deep Research augments report generation with explicit planning, iterative retrieval, evidence collection, and report updating. For completeness, the stage-specific prompt templates used in the Deep Research setting are provided in Appendix~\ref{sec:dr_prompt_templates}.

\subsection{Additional Evaluation Metrics Details}
\label{sec:app_evaluation}

We adopt Qwen3-VL-235B-A22B-Instruct because LLM-as-a-judge is a scalable protocol for open-ended generation tasks where exact-match metrics cannot capture report quality, citation quality, or evidence faithfulness.
To ensure that our conclusions do not rely on a single judge, we conduct a judge reliability study on the human-validated subset. 
The Spearman correlations with human ratings and system-level rank consistency are shown in the table.
\begin{table}[t]
\centering
\caption{Judge reliability on the human-validated subset.}
\label{tab:judge_reliability}
\small
\setlength{\tabcolsep}{2.5pt}
\begin{tabular}{lccc}
\toprule
\textbf{Dimension} & \textbf{GPT-5} &\textbf{Gemini-3.1-Pro} &\textbf{Qwen3-235B-A22B} \\
\midrule
Overall & 0.83 & 0.79 & 0.77 \\
System rank & 0.94 & 0.91 & 0.90 \\
\bottomrule
\end{tabular}
\end{table}

These results indicate that Qwen3-VL-235B-A22B-Instruct achieves great alignment with human judgments, while alternative judges produce similar system rankings, confirming that our main findings are not an artifact of a single judge model.

We further conduct robustness checks over five repeated runs with different random seeds. The results in Table~\ref{tab:repeated_runs} show consistently low standard deviations across all evaluation metrics, demonstrating the stability and reproducibility of our method.

\begin{table}[t]
\centering
\caption{
Robustness analysis over repeated runs with five random seeds.
The reported values correspond to five evaluation metrics:
Report, Traceability, Citation, Claim, Answer.
}
\label{tab:repeated_runs}
\resizebox{\linewidth}{!}{
\begin{tabular}{c|c|ccccc}
\toprule
Run & Seed & Report & Trace. & Citation & Claim & Answer \\
\midrule
Run 1 & 42   & 18.04 & 8.89 & 6.13 & 5.00 & 1.85 \\
Run 2 & 123  & 18.02 & 8.89 & 6.14 & 5.00 & 1.86 \\
Run 3 & 456  & 18.01 & 8.89 & 6.18 & 5.03 & 1.85 \\
Run 4 & 789  & 18.03 & 8.89 & 6.21 & 5.03 & 1.85 \\
Run 5 & 1024 & 18.02 & 8.89 & 6.23 & 5.03 & 1.86 \\
\midrule
Mean $\pm$ Std & -- & 18.02$\pm$0.01 & 8.89$\pm$0.00 & 6.18$\pm$0.04 & 5.02$\pm$0.02 & 1.85$\pm$0.01 \\
\bottomrule
\end{tabular}
}
\end{table}

\begin{figure*}[t!]
  \centering
  \includegraphics[height=0.96\textheight]{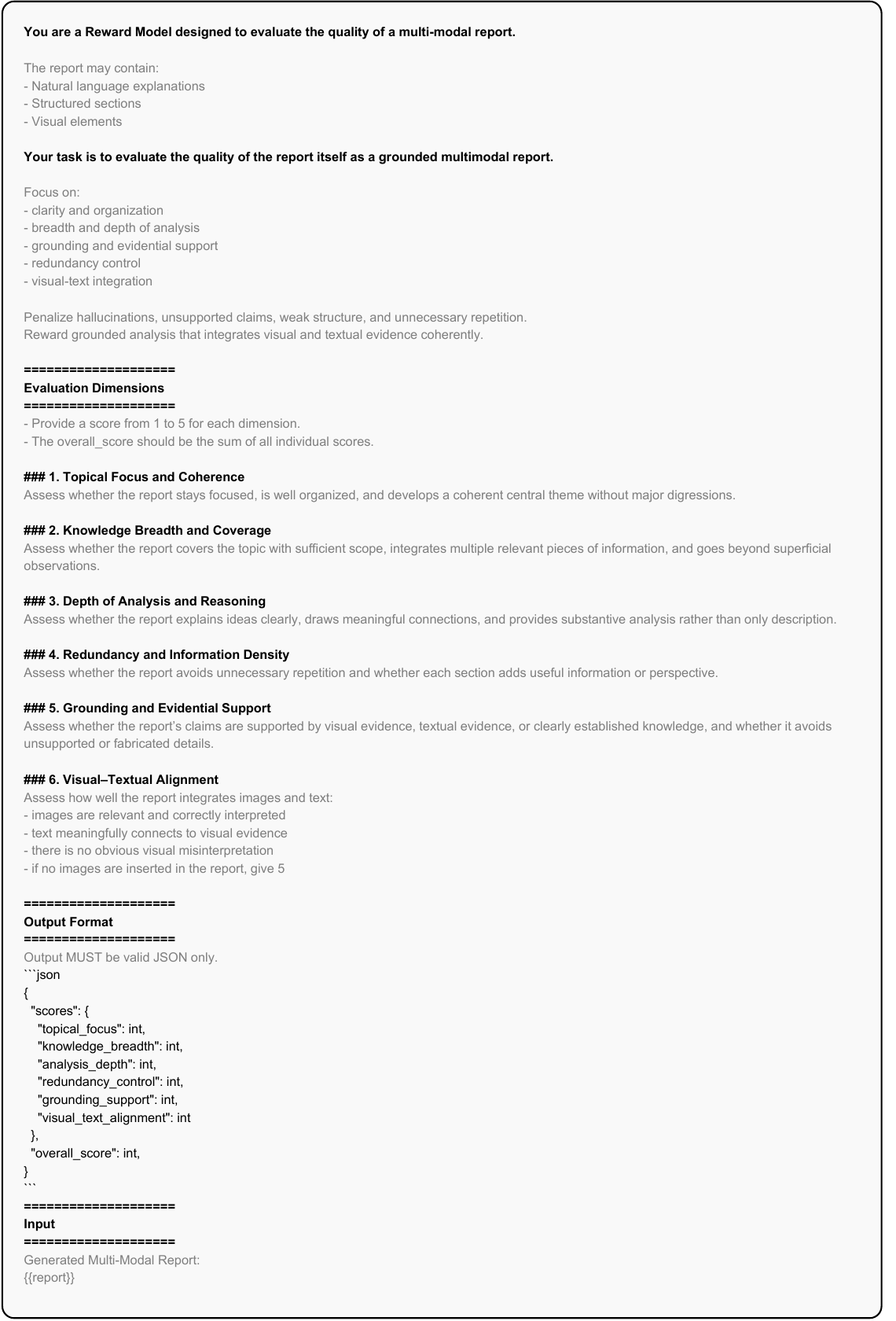}  
\caption{Prompt template used in the Evaluation of Report Dimension.}
  \label{fig:eval_r_p}
\end{figure*}

\begin{figure*}[t!]
  \centering
  \includegraphics[height=0.96\textheight]{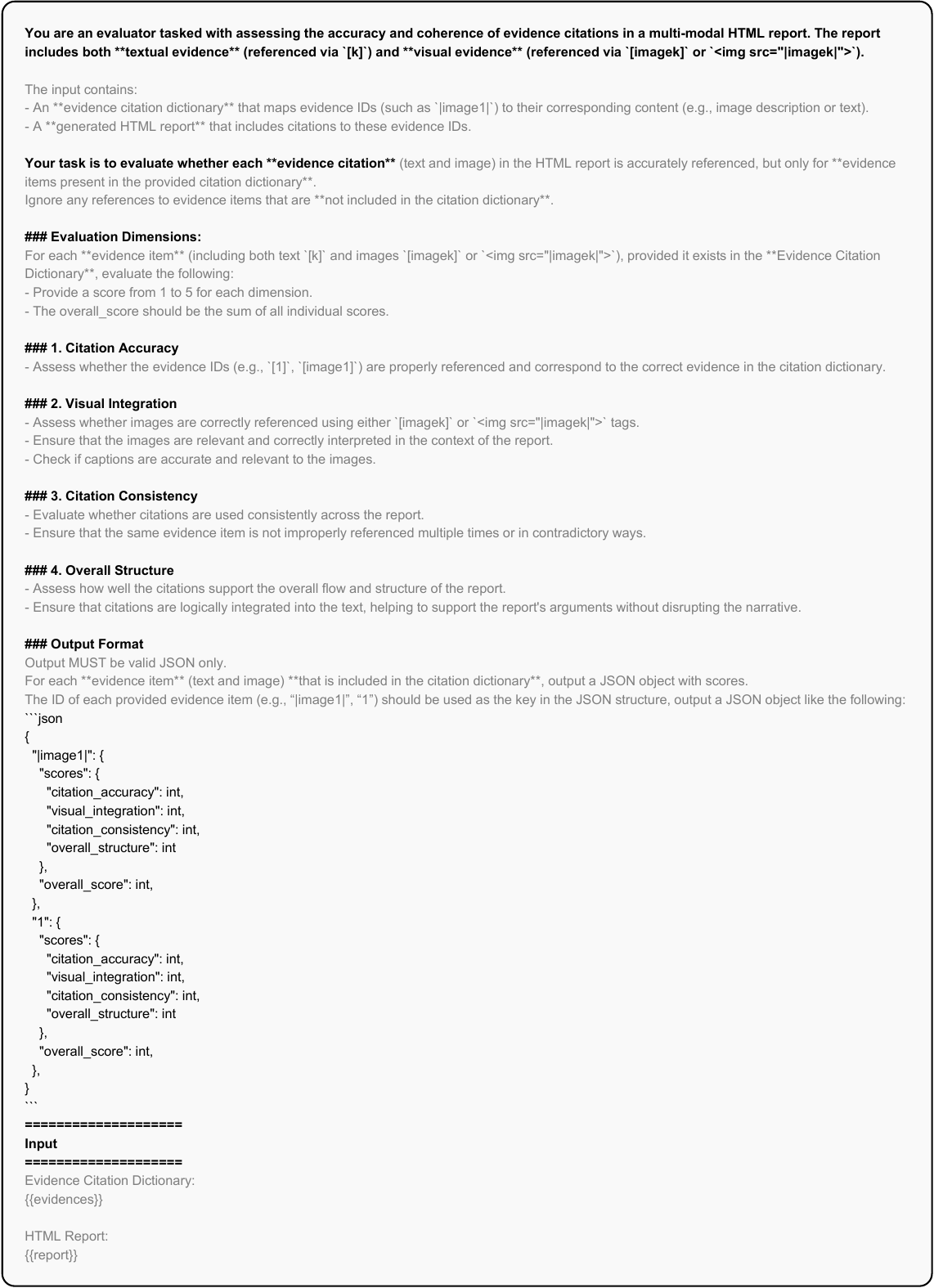}  
\caption{Prompt template used in the Evaluation of Evidence Citation Dimension.}
  \label{fig:eval_e_p}
\end{figure*}

\begin{figure*}[t!]
  \centering
  \includegraphics[height=0.96\textheight]{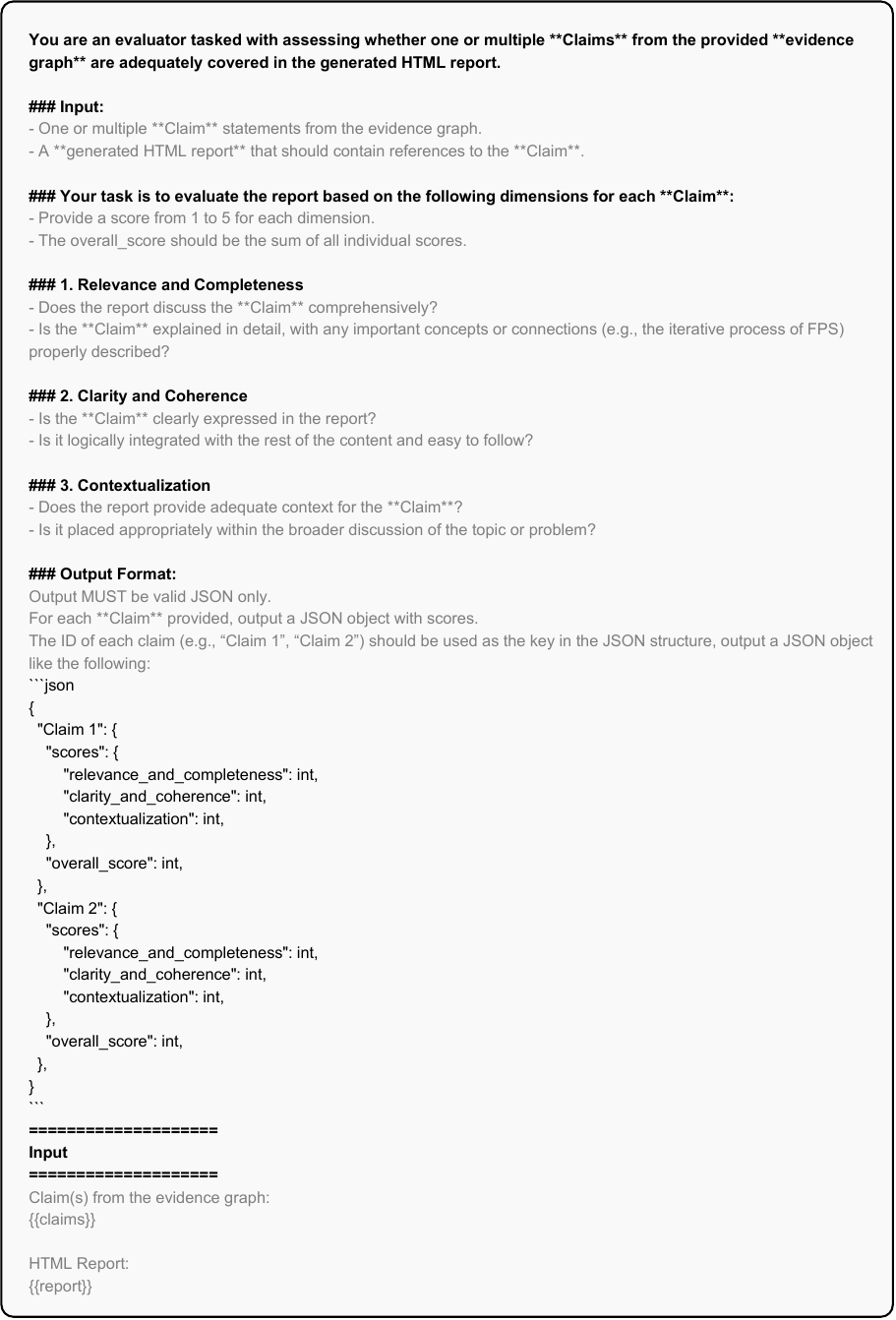}  
\caption{Prompt template used in the Evaluation of Evidence Claim Dimension.}
  \label{fig:eval_c_p}
\end{figure*}

\begin{figure*}[t!]
  \centering
  \includegraphics[height=0.96\textheight]{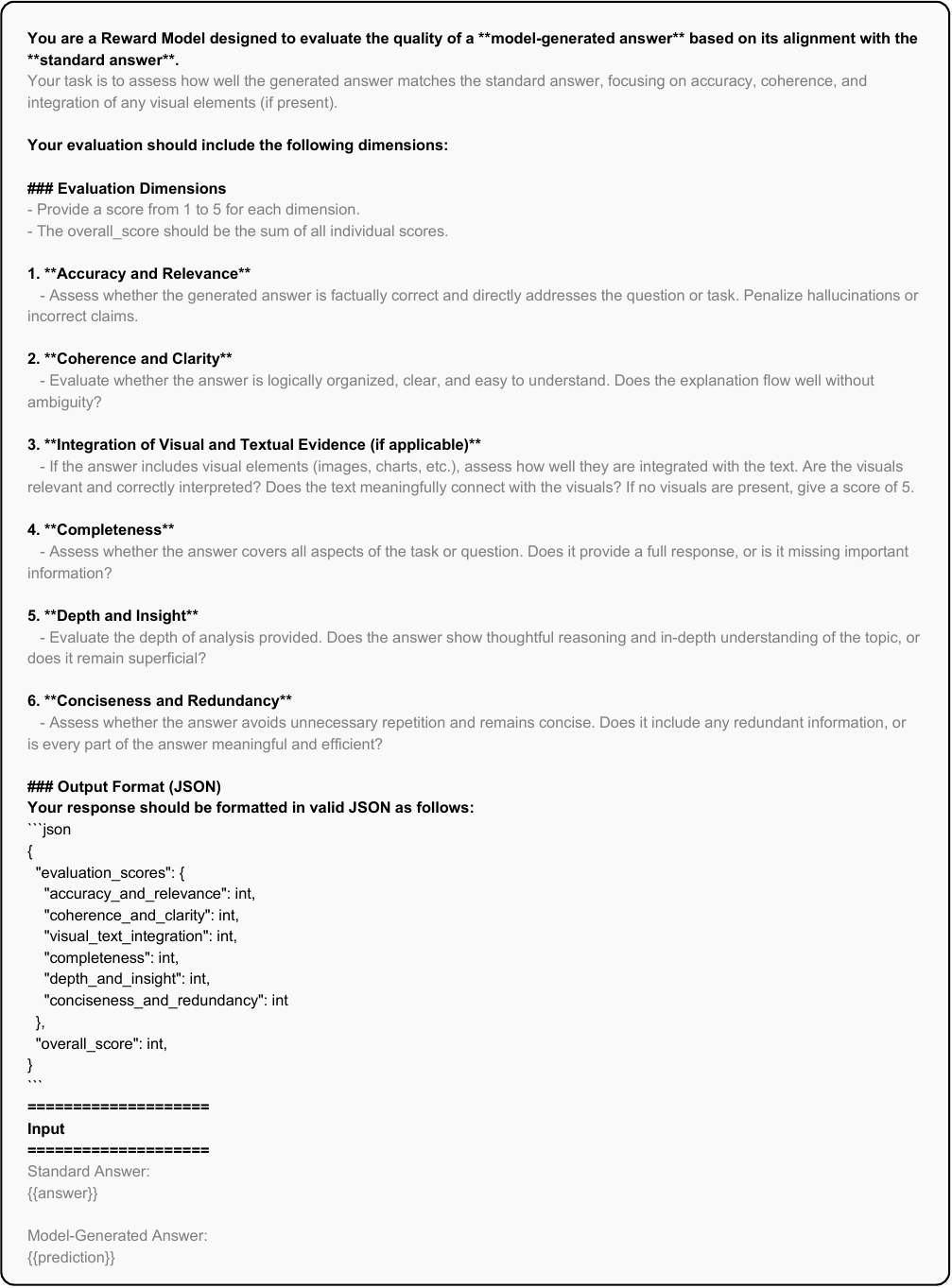}  
\caption{Prompt template used in the Evaluation of Answer Dimension.}
  \label{fig:eval_a_p}
\end{figure*}

For completeness, we further provide the prompt templates used by the LLM judges for dimension-specific evaluation. As shown in Figures~\ref{fig:eval_r_p}, \ref{fig:eval_e_p}, \ref{fig:eval_c_p}, and \ref{fig:eval_a_p}, we separately design prompts for the evaluation of the report dimension, the evidence citation dimension, the evidence claim dimension, and the answer dimension. These prompts enable fine-grained assessment of different stages of multimodal deep research, ensuring that the evaluation considers not only the overall report quality, but also the faithfulness of citation, the correctness of evidence-to-claim grounding, and the quality of the final answer. In addition, these dimension-specific evaluations are organized under our progressive gating mechanism, where a sample is evaluated at a later stage only when it passes the preceding one. By doing so, the framework avoids applying all evaluation dimensions indiscriminately to every sample, which both preserves the process-oriented nature of our benchmark for assessing intermediate evidence aggregation and reasoning quality, and improves evaluation efficiency by reducing unnecessary judge calls.

Moreover, Table~\ref{tab:rate_index_mapping} and~\ref{tab:model_index_mapping} provide the mapping between the indices used in Figure~\ref{fig:gate} and Figure~\ref{fig:evidence}, respectively.

\begin{table*}[t]
\centering
\caption{
Mapping between the indices used in Figure~\ref{fig:gate}
and the corresponding model--pipeline pairs.
Models are indexed in ascending order of their overall benchmark performance.
}
\label{tab:rate_index_mapping}
\setlength{\tabcolsep}{6pt}
\renewcommand{\arraystretch}{1.08}
\begin{tabular}{c l l c l l}
\toprule
\textbf{Index} & \textbf{Model} & \textbf{Pipeline}
&
\textbf{Index} & \textbf{Model} & \textbf{Pipeline} \\
\midrule
1  & MiniCPM-V-4.5          & RAG             &
17 & Grok-4.5              & RAG \\

2  & GLM-4.6V-Flash         & RAG             &
18 & Gemini-3.1-Flash      & RAG \\

3  & Qwen3.5-9B            & RAG             &
19 & Qwen3-VL-8B-Instruct  & Deep Research \\

4  & MiMo-VL-7B-RL         & Deep Research   &
20 & Qwen3.5-4B           & Deep Research \\

5  & Qwen3.5-4B           & RAG             &
21 & Qwen3.5-9B           & Deep Research \\

6  & MiniCPM-V-4.5         & Deep Research   &
22 & GPT-5                & RAG \\

7  & Qwen3.5-35B-A3B      & RAG             &
23 & Gemma4-31B-it        & RAG \\

8  & MiMo-VL-7B-RL         & RAG             &
24 & GPT-5-mini           & Deep Research \\

9  & GLM-4.6V-Flash        & Deep Research   &
25 & Qwen3.5-35B-A3B      & Deep Research \\

10 & Qwen3-VL-8B-Instruct  & RAG             &
26 & GPT-5.6-sol          & RAG \\

11 & Intern3.5-VL-8B       & Deep Research   &
27 & Qwen3.5-27B          & Deep Research \\

12 & Gemma4-E4B-it         & RAG             &
28 & GPT-5                & Deep Research \\

13 & Qwen3.5-27B          & RAG             &
29 & Gemini-3.1-Flash     & Deep Research \\

14 & Intern3.5-VL-8B       & RAG             &
30 & Gemma4-31B-it        & Deep Research \\

15 & GPT-5-mini           & RAG             &
31 & GPT-5.6-sol          & Deep Research \\

16 & Gemma4-E4B-it         & Deep Research   &
32 & Grok-4.5             & Deep Research \\
\bottomrule
\end{tabular}
\end{table*}

\subsection{Domain and Modality Analysis}
\label{sec:domain}

\begin{figure}[t!]
  \centering
  \includegraphics[width=\linewidth]{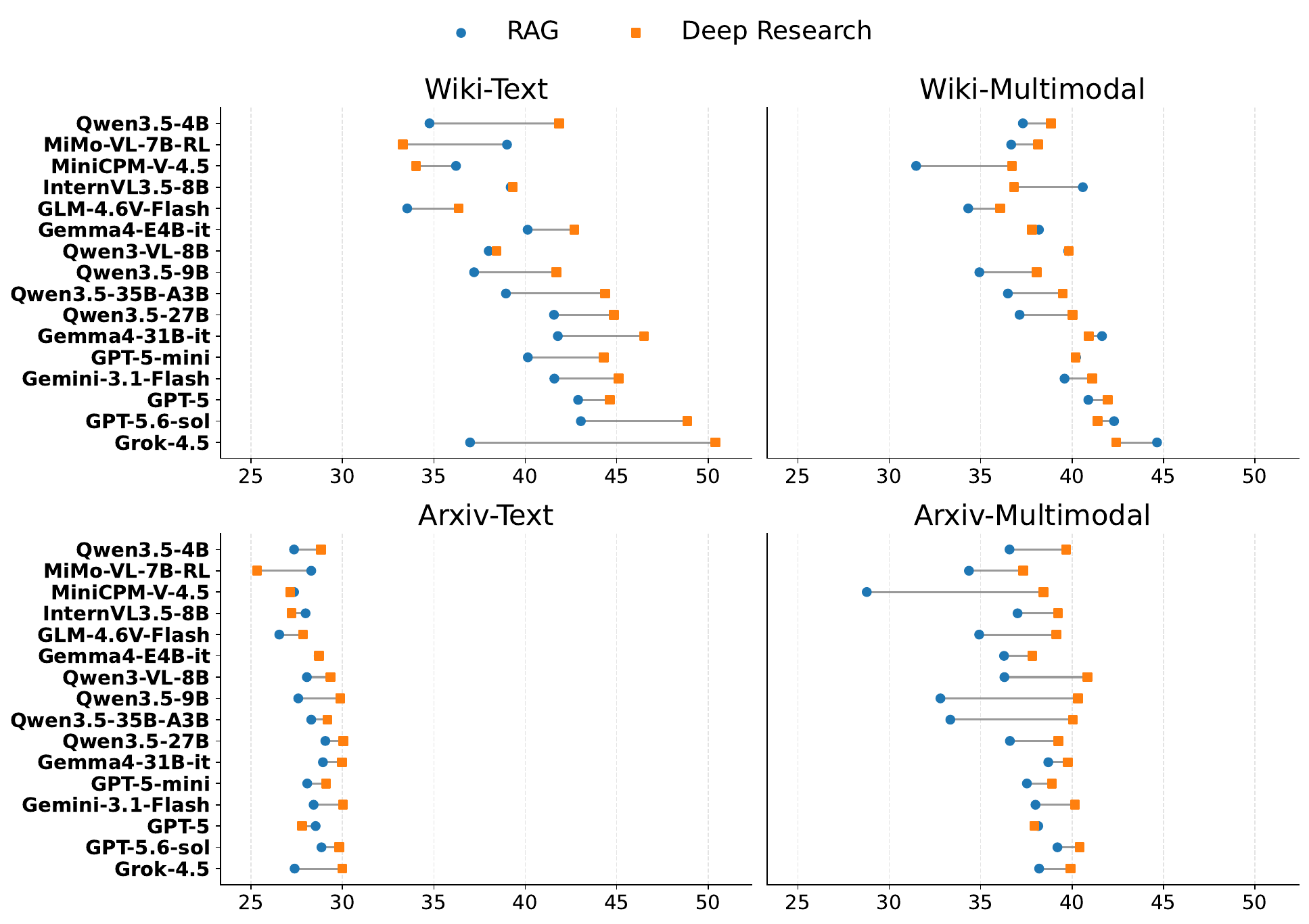}  
\caption{Comparison of RAG and Deep Research across text-only and multimodal settings in both Wiki and Arxiv subsets.}
  \label{fig:multimodal}
\end{figure}

Figure~\ref{fig:multimodal} compares RAG and Deep Research across four subsets: Wiki-Text, Wiki-Multimodal, arXiv-Text, and arXiv-Multimodal. Deep Research shows more consistent gains on the Wiki subsets, especially in the text-only setting, while its gains on the arXiv subsets are smaller and more uneven. This suggests that its effectiveness depends not only on the retrieval paradigm itself, but also on the difficulty of evidence composition in the underlying domain.

A related pattern is that multimodal settings exhibit larger variation than text-only settings in both Wiki and arXiv. This indicates that the benefits of Deep Research become less stable when stronger cross-modal grounding and more complex evidence composition are required. Taken together, these results further support our main conclusion that the key challenge of deep research lies less in evidence retrieval itself and more in organizing heterogeneous evidence into grounded intermediate claims.

\begin{table}[t]
    \centering
    \caption{
    Mapping between the model indices used in
    Figure~\ref{fig:evidence} and the corresponding model names.
    }
    \label{tab:model_index_mapping}
    \setlength{\tabcolsep}{5pt}
    \renewcommand{\arraystretch}{1.08}
    \begin{tabular}{c l c l}
        \toprule
        \textbf{Index} & \textbf{Model}
        & \textbf{Index} & \textbf{Model} \\
        \midrule
        1  & Qwen3.5-4B              & 9  & Qwen3.5-35B-A3B \\
        2  & MiMo-VL-7B-RL           & 10  & Qwen3.5-27B \\
        3  & MiniCPM-V-4.5           & 11 & Gemma4-31B-it \\
        4  & InternVL3.5-8B          & 12 & GPT-5-mini \\
        5  & GLM-4.6V-Flash          & 13 & Gemini-3.1-Flash \\
        6  & Gemma4-E4B-it           & 14 & GPT-5 \\
        7  & Qwen3-VL-8B-Instruct    & 15 & GPT-5.6-sol \\
        8  & Qwen3.5-9B              & 16 & Grok-4.5 \\
        \bottomrule
    \end{tabular}
\end{table}

\subsection{Case Study}
\label{sec:case}

\begin{figure*}[t!]
  \centering
  \includegraphics[width=1.0\textwidth]{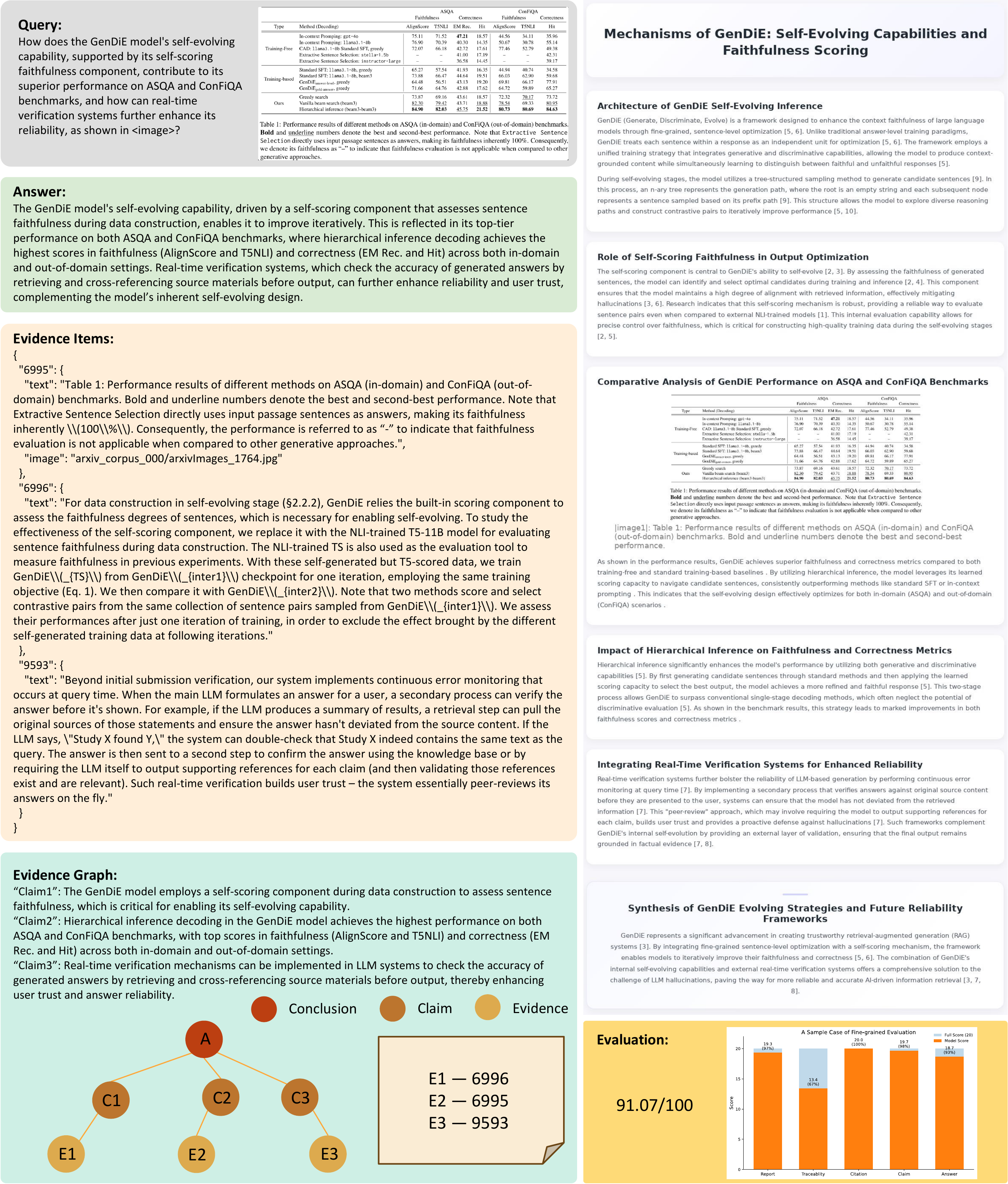}  
\caption{A case study from HiEviDR-Bench. The left panel shows the benchmark instance structure, including the question, answer, evidence items, and evidence graph; the right panel presents a rendered multimodal deep research report generated by \textit{Gemini-3.1-Flash} together with its evaluation scores.}
  \label{fig:case}
\end{figure*}

Figure~\ref{fig:case} presents a representative case from HiEviDR-Bench. Each benchmark instance is organized as a structured multimodal deep research sample, including the input question, the reference answer, a set of evidence items, and the corresponding evidence graph. The evidence items consist of retrieved multimodal materials associated with unique identifiers, while the evidence graph further organizes them into a hierarchical reasoning structure that links evidence nodes to intermediate claim nodes and finally to the conclusion node. In this way, the benchmark does not only specify what the final answer should be, but also explicitly models how supporting evidence should be aggregated and connected during the reasoning process.

The right part of Figure~\ref{fig:case} shows an example report generated by \textit{Gemini-3.1-Flash-Lite-Preview} under the Deep Research setting. The generated report is rendered in a multimodal format, where textual analysis is accompanied by properly inserted visual evidence at suitable positions. Compared with a purely textual report, this multimodal presentation can convey richer information by directly exposing relevant tables, figures, and other visual materials together with the associated explanation, thereby making the report more informative and easier to inspect. This example also illustrates the practical advantage of our report generation framework, which supports the integration of retrieved images into the final rendered output rather than limiting the report to text-only summaries.

The bottom-right panel further shows the fine-grained evaluation result for this case. Under our progressive gating mechanism, the sample successfully passes all evaluation stages, indicating that the model not only produces a high-quality report, but also cites evidence appropriately, establishes valid evidence-to-claim grounding, and arrives at a well-supported final answer. Correspondingly, the model achieves consistently strong scores across all five dimensions, demonstrating that it can generate a content-rich and well-grounded multimodal deep research report for this case.

\subsection{Prompt Templates for Deep Research}
\label{sec:dr_prompt_templates}

For completeness, we provide the stage-specific prompt templates used in the Deep Research setting in this subsection. These prompts correspond to the three main stages of the pipeline, namely planning, initialization, and iterative updating. Together, they govern how the model decomposes the research task, organizes retrieved evidence, and progressively refines the report during the multi-step generation process.

\begin{figure*}[t!]
  \centering
  \includegraphics[height=0.96\textheight]{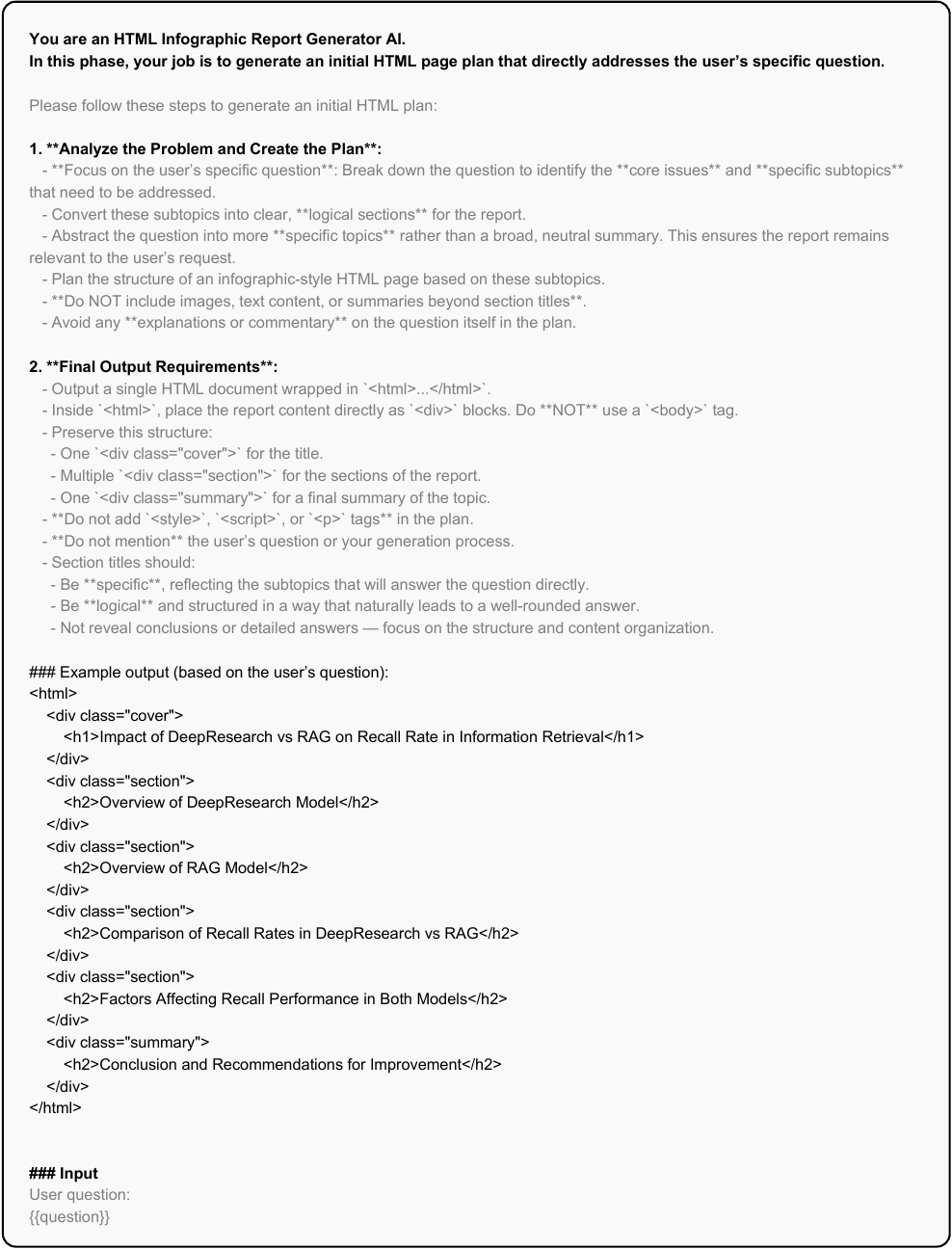}  
\caption{Prompt template used in the Deep Research Plan Stage.}
  \label{fig:dr_plan_p}
\end{figure*}

\begin{figure*}[t!]
  \centering
  \includegraphics[height=0.96\textheight]{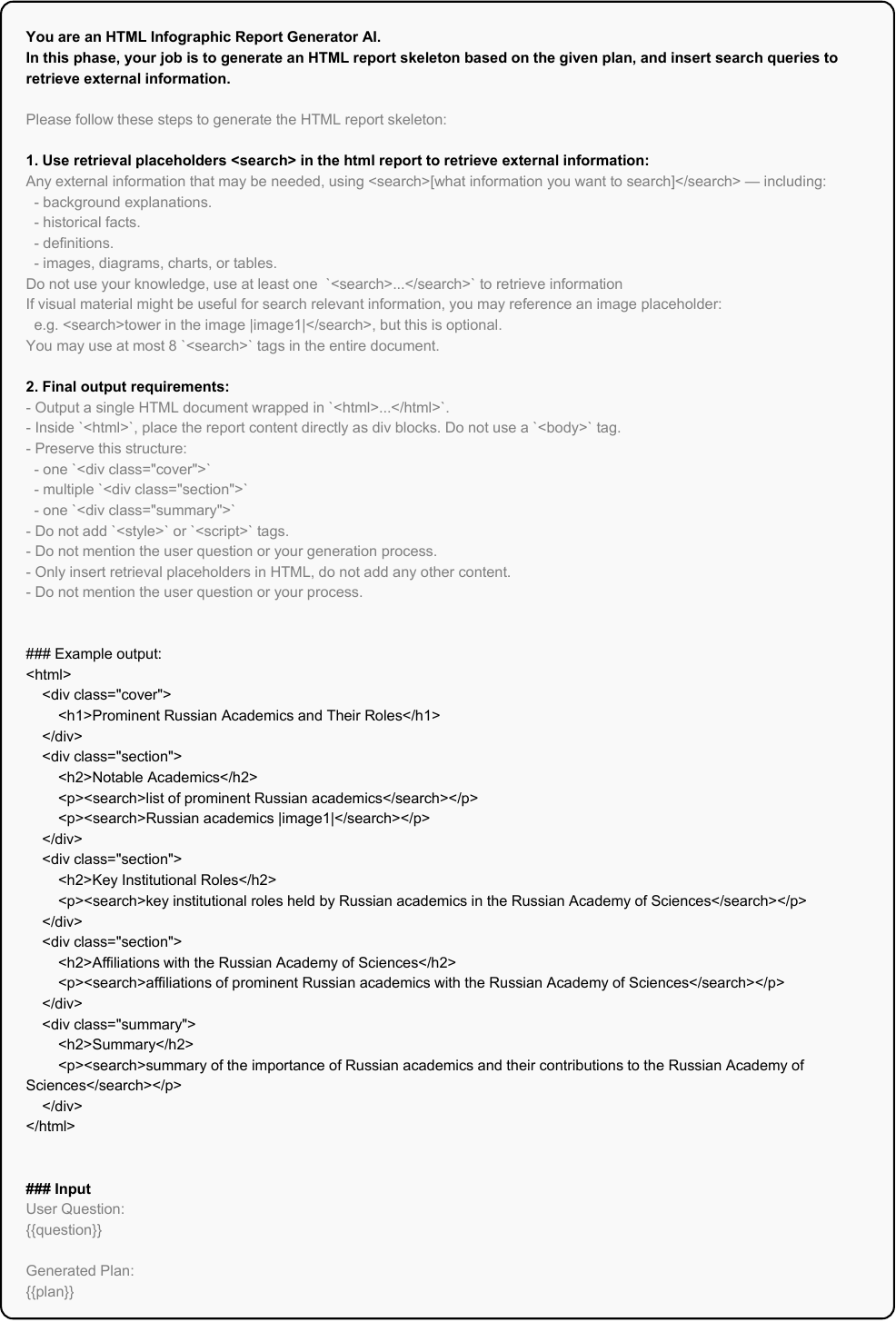}  
\caption{Prompt template used in the Deep Research Init Stage.}
  \label{fig:dr_init_p}
\end{figure*}

\begin{figure*}[t!]
  \centering
  \includegraphics[height=0.9\textheight]{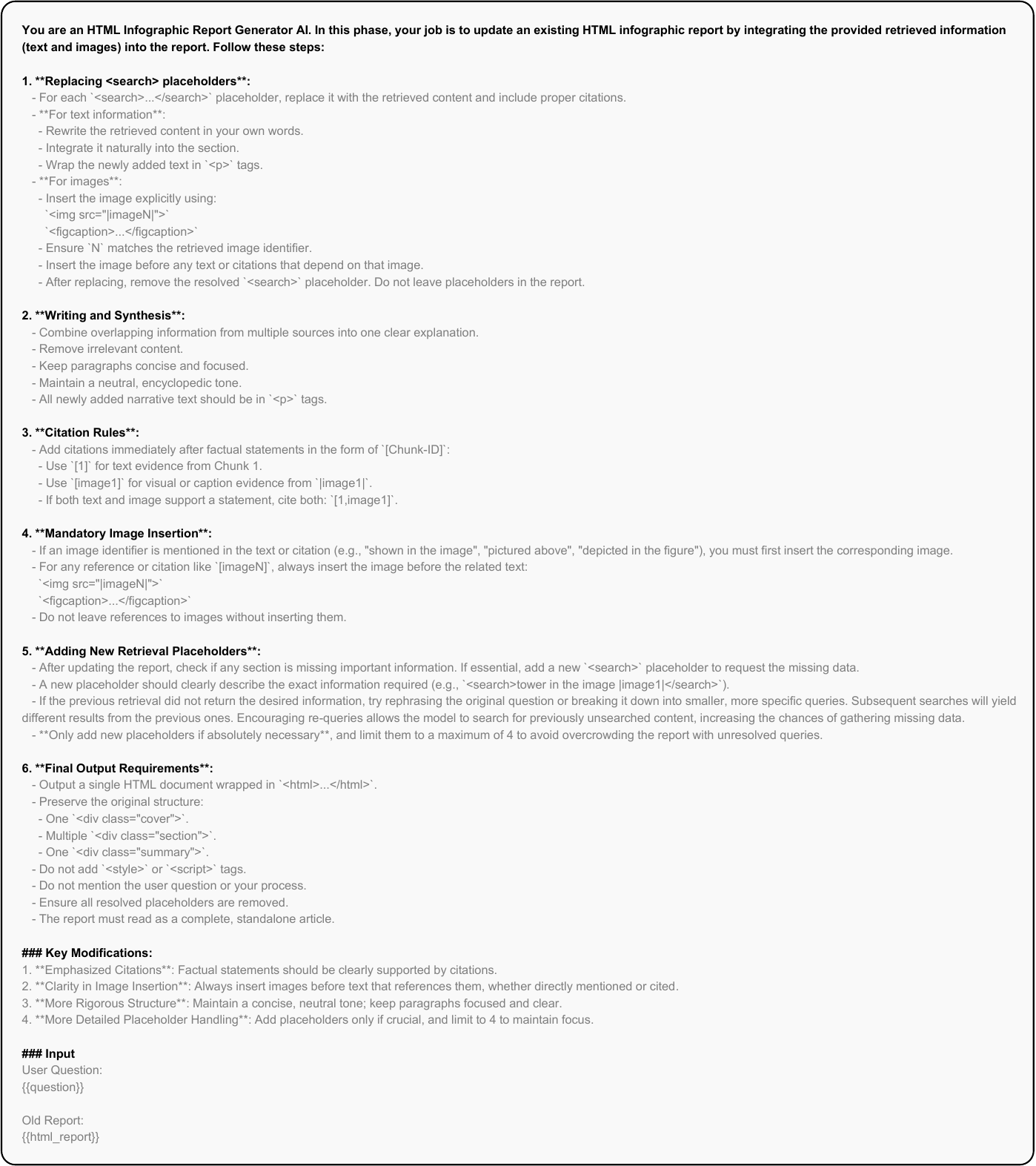}  
\caption{Prompt template used in the Deep Research Update Stage.}
  \label{fig:dr_update_p}
\end{figure*}

\end{document}